\documentclass[iop]{emulateapj}
\newcommand{\noprint}[1]{}

\usepackage{graphicx}

\newcommand{\nthp}{N$_2$H$^+$}
\newcommand{\amm}{NH$_3$}
\newcommand{\tcts}{$^{13}$C$^{34}$S}
\newcommand{\tcs}{$^{13}$CS}
\newcommand{\cth}{C$_{2}$H}
\newcommand{\chtcn}{CH$_{3}$CN}
\newcommand{\htcn}{H$^{13}$CN}
\newcommand{\htcop}{H$^{13}$CO$^{+}$}
\newcommand{\hfa}{H41$\alpha$}
\newcommand{\hctn}{HC$_{3}$N}
\newcommand{\hctccn}{HC$^{13}$CCN}
\newcommand{\hcn}{HCN}
\newcommand{\hnc}{HNC}
\newcommand{\hcop}{HCO\ensuremath{^{+}}}
\newcommand{\hncofzf}{HNCO 4$_{0,4}$}
\newcommand{\hncofot}{HNCO 4$_{1,3}$}
\newcommand{\sio}{SiO}
\newcommand{\hnco}{HNCO}

\newcommand{\Msun}{\ensuremath{M_{\odot}}}

\newcommand{\HII}{H II}
\newcommand{\FWHM}{\ensuremath{\Delta V_{\textrm{FWHM}}}}

\shorttitle{MALT90 Pilot Survey}
\shortauthors{Foster}
\begin{document}
\title{The Millimeter Astronomy Legacy Team 90 GHz (MALT90) Pilot Survey}

\author{Jonathan B. Foster\altaffilmark{1}, James M. Jackson\altaffilmark{1}, Elizabeth Barris\altaffilmark{1}, Kate Brooks\altaffilmark{2}, Maria Cunningham\altaffilmark{3}, Susanna C. Finn\altaffilmark{1}, Gary A. Fuller\altaffilmark{4}, Steve N. Longmore\altaffilmark{5,6}, Joshua L. Mascoop\altaffilmark{1}, Nicolas Peretto\altaffilmark{7}, Jill Rathborne\altaffilmark{2,8}, Patricio Sanhueza\altaffilmark{1}, Fr\'ed\'eric Schuller\altaffilmark{9}, Friedrich Wyrowski\altaffilmark{9}}

\email{jbfoster@bu.edu}

\altaffiltext{1}{Institute for Astrophysical Research, Boston University, Boston, MA 02215, USA}
\altaffiltext{2}{CSIRO Astronomy and Space Science, PO Box 76, Epping, NSW 1710, Australia}
\altaffiltext{3}{School of Physics, The University of New South Wales, Sydney 2052, Australia}
\altaffiltext{4}{Jodrell Bank Centre for Astrophysics, School of Physics and Astronomy, \\ University of Manchester, Manchester M13 9PL, UK}
\altaffiltext{5}{ESO Headquarters, Karl-Schwarzschild-Str. 2, 85748 Garching bei M\"{u}nchen, Germany}
\altaffiltext{6}{Harvard-Smithsonian Center for Astrophysics, MS 42, 60 Garden Street, \\ Cambridge, MA 02138, USA}
\altaffiltext{7}{Laboratoire AIM, CEA/DSM-CNRS-Universit\^{e} Paris Diderot, IRFU/Service d'Astrophysique, C.E. Saclay, Orme de Merisiers, 91191 Gif-sur-Yvette, France}
\altaffiltext{8}{Departamento de Astronom\'ia, Universidad de Chile, Chile}
\altaffiltext{9}{Max-Plank-Institut f\"{u}r Radioastronomie, Auf dem H\"{u}gel 69, D-53121 Bonn, Germany}

\begin{abstract}
We describe a pilot survey conducted with the Mopra 22-m radio telescope in preparation for the Millimeter Astronomy Legacy Team Survey at 90 GHz (MALT90). We identified 182 candidate dense molecular clumps using six different selection criteria and mapped each source simultaneously in 16 different lines near 90 GHz.  We present a summary of the data and describe how the results of the pilot survey shaped the design of the larger MALT90 survey. We motivate our selection of target sources for the main survey based on the pilot detection rates and demonstrate the value of mapping in multiple lines simultaneously at high spectral resolution.
\end{abstract}


\section{Introduction}

The goal of the Millimeter Astronomy Legacy Team Survey at 90 GHz (MALT90) is to characterize the physical and chemical conditions of dense molecular clumps associated with high-mass star formation over a wide range of evolutionary states. MALT90 will do this by taking advantage of the newly upgraded Mopra Spectrometer (MOPS\footnotemark[1]) and the fast mapping capability of the Mopra 22-m radio telescope\footnotemark[2]. The survey will obtain molecular line maps of 3000 candidate dense molecular clumps. The clumps will be selected so as to cover a broad range of evolutionary states, from pre-stellar clumps to accreting high-mass protostars and on to \HII\ regions. The survey will be conducted at 90 GHz because this frequency regime contains numerous molecular lines which have typical critical densities for collisional excitation of $ \gtrsim 10^5$ cm$^{-3}$ and are therefore excellent tracers of dense gas. Such data will allow us to study the Galactic distribution of these clumps, their physical properties, and their chemical variation and evolution; this basic information is necessary to constrain theories of high-mass star formation. In addition, MALT90 will provide a valuable database of dense molecular clumps associated with high-mass star formation for future ALMA observations.

MALT90 will map roughly 3,000 dense molecular clumps, providing an order of magnitude more sources than previous comparable surveys \citep[e.g.,][]{Shirley:2003, Pirogov:2003, Gibson:2009, Wu:2010}. A large number of sources will allow us to divide the sample into sub-samples (based on mass, evolutionary phase, etc.) yet retain a sufficient number of sources in each sub-sample for statistical analysis. Because dense molecular gas occupies only a small solid angle of the Galactic plane and molecular emission at 90 GHz is relatively faint, a blind fully-sampled 90 GHz survey of a significant portion of the Galactic plane is impractical. Instead, we must choose targets based on other methods for identifying dense molecular clumps. The main purpose of the MALT90 pilot survey described herein is to choose the best method for identifying dense molecular clumps, with the twin aims of having a high percentage of detections within our sensitivity limits and covering a broad range of evolutionary states. 

Throughout this paper, we will use the term ``dense molecular clump'' to refer to our sources. The choice of ``clump'' follows the naming system used by \citet{Williams:2000} and \citet{Bergin:2007} which distinguishes between molecular clouds, clumps, and cores. In this scheme, clumps are coherent regions in position-velocity space with typical masses of 50-500 \Msun, typical sizes of 0.3-3 pc and typical mean densities of 10$^3$-10$^4$ cm$^{-3}$ which may contain additional substructures called cores which give rise to individual stars or stellar systems. Our goal in MALT90 is to identify and map the clumps that give rise to a cluster of stars containing one or more high-mass stars.

This paper will focus predominantly on the technical validation of the survey and explain the design choices motivated by the pilot survey. Data from the pilot survey will be combined with the full MALT90 survey data (where the diverse selection criteria in the pilot survey are not detrimental to statistical analysis) for the specific scientific projects in MALT90. These analyses will appear in future papers.  

\footnotetext[1]{MOPS was funded in part through a grant instigated by the University of New South Wales (UNSW) under the Australian Research Council Grants scheme for Linkage, Infrastructure, Equipment and Facilities (LIEF), and in part by CSIRO Astronomy and Space Science.}

\footnotetext[2]{The Mopra radio telescope is part of the Australia Telescope National Facility which is funded by the Commonwealth of Australia for operation as a National Facility managed by CSIRO.}

\section{Target Selection}

We used six different input catalogues for selecting sources; three were lists that we produced for our pilot survey and three were based on pre-existing catalogs. From these lists we chose 20 - 40 sources near integer Galactic longitudes covering the range of longitudes accessible from Mopra. This assured a random selection of sources from the three pre-compiled lists and allowed us to focus on a limited portion of the sky when developing our own lists, while still covering a broad range of Galactic longitudes. The six selection criteria are summarized in Table~\ref{surveys}.

\begin{deluxetable*}{llll}
\tablecaption{MALT90 Pilot Survey Input Catalogs}
	
	\tablehead{\colhead{Data} & \colhead{Criterion} & \colhead{Object Identified} & \colhead{Shorthand}}
	\startdata
	GLIMPSE 3.6 to 8~\micron		&		Dark Extinction			&	Pre-stellar (IRDC) Clump		& GLIMPSE-Dark	\\
	GLIMPSE 8~\micron				&		Extended Emission		&	\HII\ Region				& GLIMPSE-Bright	\\
	MIPSGAL 24~\micron			&		Point Source			&	Accreting Protostar			& MIPSGAL	\\
	HOPS\tablenotemark{a} 1.2 cm 				&		Compact \amm\ Source	&	Dense Clump				& HOPS	\\
	IRAS +  1.2 mm	 emission\tablenotemark{b}				&		IRAS + mm Continuum	&	Star-forming Dense Clump			& IRAS	\\
	ATLASGAL\tablenotemark{c} 870~\micron		&		Compact Continuum		&	Dense Clump				& ATLASGAL	\\
	\enddata
	\tablenotetext{a}{\citet{Walsh:2008}}
	\tablenotetext{b}{\citet{Beltran:2006}}
	\tablenotetext{c}{\citet{Schuller:2009}}
\label{surveys}
\end{deluxetable*}

\begin{figure}
\epsscale{1.15}
\plotone{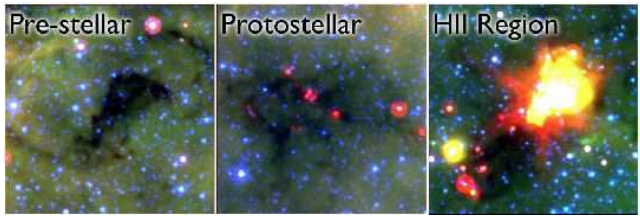}
\caption{A proposed evolutionary sequence for high-mass star formation, running from pre-stellar clumps [left], to clumps with embedded accreting protostars [middle], to \HII\ regions [right]. These three-color images show IRAC band 1 (3.6 \micron) as blue, IRAC band 4 (8.0~\micron) as green, and MIPS 24~\micron\ as red. These categories correspond to different \emph{Spitzer} emission morphologies and motivated the three different catalogs based on GLIMPSE and MIPSGAL data. Specifically, clumps which are dark at GLIMPSE wavelengths (3.6 - 8~\micron) are identified as pre-stellar, clumps with a MIPSGAL 24~\micron\ point source are identified as protostellar, and clumps with extended 8~\micron\ flux are identified as \HII\ regions.}
\label{spitzer}
\end{figure}

The first three lists were produced for our pilot survey. For these lists we chose sources based on the mid-infared morphology of candidate dense molecular clumps as revealed by two \emph{Spitzer} Space Telescope Legacy surveys: the Galactic Legacy Infrared Mid-Plane Survey Extraordinaire \citep[GLIMPSE;][]{Benjamin:2003} and the 24 and 70 Micron Survey of the Inner Galactic Disk with MIPS \citep[MIPSGAL;][]{Carey:2009}. GLIMPSE covers the Galactic plane in the Infrared Array Camera \citep[IRAC;][]{Fazio:2004} bands from 3.6 to 8.0~\micron, while MIPSGAL covers much the same area at 24 and 70~\micron\ using the Multiband Infrared Photometer for \emph{Spitzer} \citep[MIPS;][]{Rieke:2004}. 

We examined the GLIMPSE and MIPSGAL mosaics using three criteria designed to select sources in distinct evolutionary states as shown in Figure~\ref{spitzer}. A preliminary version of the \citet{Peretto:2009} catalog of Infrared Dark Clouds (IRDCs) was used to identify 8~\micron\ extinction features near integer Galactic longitudes. This catalog was then trimmed to remove any sources which contained a 24~\micron\ point source (since this is most likely a protostar). We refer to this the GLIMPSE-Dark catalog, and these sources should correspond to the earliest phase of high-mass star formation. We then examined the \emph{Spitzer} mosaics near integer Galactic longitudes by hand to choose candidate dense molecular clumps containing either 24~\micron\ point sources or bright extended 8~\micron\ emission. If a clump of 8~\micron\ dark extinction was coincident with a 24~\micron\ point source, we assigned it to the MIPSGAL catalog and classified this candidate clump as protostellar. In the case of bright extended emission at 8~\micron\ we assigned the source to the GLIMPSE-Bright catalog and classified it as an \HII\ region.

The correspondence between the appearance of a source in the \emph{Spitzer} surveys and its evolutionary state is clearly imperfect. The projection of unrelated objects along a given line of sight, inhomogeneities in the diffuse 8~\micron\ emission, and sensitivity limits (e.g., our ability to detect 24~\micron\ point sources will depend on intrinsic luminosity and distance) are three possible sources of misidentification. However, this system provides a quick and uniform way to make an initial assessment of a candidate dense molecular clump's evolutionary state.

The other three lists in our pilot survey were produced using pre-existing catalogs. The first came from the H$_2$O Southern Galactic Plane Survey \citep[HOPS;][]{Walsh:2008} which used Mopra at 1.2 cm to map the Galactic plane from $-70^\circ < l < 30^\circ$ in \amm\ and H$_2$O. Because the HOPS \amm\ (1,1) and (2,2) lines have a similar critical density (n $\sim$ 10$^{4}$ cm$^{-3}$) as the MALT90 90 GHz lines (n $\sim$ 10$^{5}$ cm$^{-3}$), bright \amm\ sources in HOPS are likely to be detected by Mopra in the 90 GHz lines. The last two catalogs came from mm/submm continuum surveys which reveal the location of regions with high dust column density, typically corresponding to dense molecular clumps. The \citet{Beltran:2006} survey at 1.2 mm made maps around Infrared Astronomical Satellite (IRAS) point sources using the Swedish-ESO SubMillimeter Telescope (SEST) and the SEST Imaging Bolometer Array (SIMBA). Because the \citet{Beltran:2006} maps were made toward IRAS point sources, these sources are likely to contain a protostar or \HII\ region, which gives rise to the IRAS emission; we shall refer to this catalog as the IRAS catalog for convenience. Finally, the APEX (Atacama Pathfinder Experiment) Telescope Large Area Survey of the Galaxy \citep[ATLASGAL;][]{Schuller:2009} is a survey of the Galactic plane ($\pm$ 60\arcdeg\ in longitude over $\pm$ 1.5\arcdeg\ in latitude and -80\arcdeg $\leq$ l $\leq$ -60\arcdeg\ with -2\arcdeg $\leq$ b $\leq$ 1\arcdeg) at 870~\micron. From a preliminary compact source catalog we chose ATLASGAL sources with peak fluxes above 2 Jy/beam closest to integer Galactic longitudes.

In two cases (G336.994$-$00.019 and G339.968$-$00.529), a candidate dense molecular clump was chosen independently from two different catalogs. We consider these sources to belong to both catalogs when considering detection statistics. 

In summary, we have combined six separate lists, four of which we expected to select particular evolutionary states (the \emph{Spitzer}-identified sources and the \citet{Beltran:2006} catalog of IRAS sources), and two of which (ATLASGAL and HOPS) we expected to be less biased with respect to evolutionary status. For both ATLASGAL and HOPS, the most luminous sources will tend to be the hottest, more evolved sources. To select a broad range of evolutionary states from these surveys it is necessary to include some additional information as we discuss in \S~\ref{Choice}. 

\section{Data}

We carried out observations for the MALT90 pilot survey in the austral winter of 2009 from June 15-24. The On-The-Fly (OTF) mapping mode of Mopra was used. Maps were made with the beam center running on a 3\arcmin.4 x 3\arcmin.4  grid. At typical distances to high-mass star-forming regions (several kpc) this map size is sufficient to cover the expected spatial extent of a few parsecs for our dense molecular clumps. The scan rate was 3.92\arcsec\ per second. The map is made with 12\arcsec\ spacing between the rows, giving 17 rows per map. Since the Mopra beam at 90 GHz is 36\arcsec, this row spacing provides redundancy in the map. OFF positions were chosen at $\pm$ 1 degree in Galactic latitude away from the plane (positive offset for sources at positive Galactic latitude and vice-versa), and though the OFF positions were not explicitly checked for line emission, no map showed evidence of contamination from signal in the OFF. A single OFF position was observed for every two scan rows. In general, maps were made scanning in strips of constant Galactic longitude, although for two sources maps were also taken by scanning in strips of constant Galactic latitude.

Pointing on SiO masers was performed every 1-1.5 hour, maintaining pointing precision to better than about 10\arcsec. Typical system temperatures (T$_{sys}$) were 150 - 250 K and were measured by paddle scans every 15 minutes. Weather conditions were variable. Sources observed under poor system temperatures (T$_{sys}$ $>$ 500 K) or rapidly varying conditions were re-observed under more clement conditions. For each source, only the map taken with the lowest T$_{sys}$ is presented here.

The full 8 GHz bandwidth of MOPS was split into 16 zoom bands of 138 MHz each providing a velocity resolution of $\sim$ 0.11 km s$^{-1}$ in each band, easily sufficient to resolve line emission in a high-mass star-forming region. The central frequencies are shown in Table~\ref{lines}, along with the line targeted at that frequency and what information that line primarily provides. The strongest lines were \nthp (1-0), \hnc (1-0), \hcop (1-0), and \hcn (1-0). These lines are all good tracers of dense gas, but provide slightly different information. \nthp\ is more resistant to freeze-out on grains than the carbon-bearing species \citep[][]{Bergin:2001}. \hnc\ is particularly prevalent in cold gas \citep[][]{Hirota:1998}. \hcop\ often shows infall signatures and outflow wings \citep[e.g.,][]{Rawlings:2004, Fuller:2005}. These strong lines can all be optically thick. Two isotopologues, \htcop\ (1-0) and \htcn\ (1-0) were also observed and provide optical depth and line profile information. \tcs\ (2-1) is another optically thin column density tracer by virtue of its low abundance. We also include \tcts\ (2-1) but this molecule is too rare to be detected.

A number of lines were chosen as tracers of hot core chemistry: \chtcn\ ($J_K = 5_1 - 4_1$), \hctn\ ($J = 10 - 9 $), \hctccn\ ($J = 10 - 9, F = 9 - 8$), \hnco\ ($J_{K_a,K_b}=4_{0,4}-3_{0,3}$), \hnco\ ($J_{K_a,K_b}=4_{1,3}-3_{1,2}$) \citep[][]{Brown:1988}. These carbon-bearing species are typically only seen in the hot cores around high-mass protostars once molecules have been liberated off dust grains by radiation or shocks. Three more lines trace particular environments:  the recombination line \hfa\ traces ionized gas \citep{Shukla:2004}; \sio\ (2-1) is seen when \sio\ is formed from shocked dust grains, typically in outflows \citep{Schilke:1997}; \cth\ is produced in photodissociation regions \citep[e.g.,][]{Lo:2009, Gerin:2011}, the $N = 1 - 0, J = 3/2 - 1/2, F =  2 - 1 $ transition is the strongest of several \cth\ lines in this spectral window. Henceforth we will refer to these line transitions by the molecule name where this usage is unambiguous (i.e. \hcop\ instead of \hcop (1-0)).

The maps were reduced using the \textsc{Livedata} and \textsc{Gridzilla} packages\footnote{https://www.atnf.csiro.au/people/mcalabre/livedata.html}. \textsc{Livedata} performs bandpass calibration using reference OFF scans and fits a 2nd order polynomial to the baseline. \textsc{Gridzilla} uses this output to construct a uniformly gridded cube. Both polarizations were averaged together. A top-hat smoothing kernel with radius of 30\arcsec\ was used to determine which spectra contribute signal to a pixel in the output map, and spectra were weighted by the system temperature. This choice of parameters produces an effective beam size of FWHM = 72\arcsec. The final cube is over-sampled in spatial frequency (9\arcsec\ pixels) and is  4\arcmin.6 x 4\arcmin.6 with the edges having significantly lower coverage, i.e. integration time. The data are presented on the antenna temperature scale (T$_{A}^{*}$). The beam efficiency of Mopra is between 0.49 at 86 GHz and 0.42 at 115 GHz \citep{Ladd:2005} for converting T$_{A}^{*}$ into main-beam brightness temperature (T$_{mb}$). All the data are publicly available from the MALT90 website\footnote{http://malt90.bu.edu}.

\begin{deluxetable*}{ccccl}
\tablecaption{MALT90 Pilot Survey Lines}
	\tablehead{\colhead{IF} & \colhead{Species} & \colhead{Transition} & \colhead{$\nu_{0}$ (GHz)} & \colhead{Primary Information Provided}}
	\startdata
1	&	\nthp		&	$J = 1 - 0$						& 93.173772	&	High column density, depletion resistant, optical depth	\\
2	&	\tcs		&	$J = 2 - 1$         						& 92.494303	&	High column density			\\
3	&	H		&	41$\alpha$       						& 92.034475	&	Ionized gas			\\
4	&	\chtcn	&	$J_K = 5_1 - 4_1$					& 90.979020	&	Hot core			\\
5	&	\hctn		&	$J = 10 - 9 $						& 91.199796	&	Hot core			\\
6	&	\tcts		&	$J = 2 - 1$              					& 90.926036	&	High colum density			\\
7	&	\hnc		&	$J = 1 - 0$						& 90.663572	&	High column density, cold gas		\\
8	&	\hctccn	&	$J = 10 - 9, F = 9 - 8$				& 90.593059	&	Hot core			\\
9	&	\hcop	&	$J = 1 - 0$						& 89.188526	&	High column density, kinematics		\\
10	&	\hcn		&	$J = 1 - 0$						& 88.631847	&	High column density, optical depth			\\
11	&	HNCO	&	$J_{K_a,K_b}=4_{0,4}-3_{0,3}$		& 88.239027	&	Hot core			\\
12	&	HNCO	&	$J_{K_a,K_b}=4_{1,3}-3_{1,2}$		& 87.925238	&	Hot core			\\
13	&	\cth		&	$N = 1 - 0, J = 3/2 - 1/2, F =  2 - 1 $		& 87.316925	&	Photodissociation region	\\
14	&	\sio		&	$J = 2 - 1$						& 86.847010	&	Shock/outflow			\\
15	&	\htcop	&	$J = 1 - 0$						& 86.754330	&	High column density, optical depth		\\
16	&	\htcn		&	$J = 1 - 0$						& 86.340167	&	High column density, optical depth		\\
\enddata
\label{lines}
\tablecomments{Frequencies listed above are the rest frequencies used in converting to velocity scale (FITS keyword RESTFREQ). }
\end{deluxetable*}

\section{Analysis}
\label{analysis}
The MALT90 pilot survey data were used to test some of the analysis tools in development for the full survey. Three different methods were used to assess detection statistics: (1) ``by-hand'' examination, (2) automated Gaussian fitting based on the HOPS \citep{Walsh:2008} pipeline and (3) moment maps. Although Gaussian fitting is critical for certain measurements (particularly for lines with hyperfine component), moment maps are a fast and relatively robust way to measure basic line properties and this study will focus only on properties well-measured by moment analysis. The results of the ``by-hand'' examination were used to select the parameters used in generating moment maps. 

As a first step to making moment maps, we calculated an error map for each spatial pixel by computing the standard deviation of the spectra at that position using 3$\sigma$ iterative rejection. In this way, we remove strong line features from the calculation and derive an estimate of the per-channel noise in the spectrum at each point in the map. Our on-the-fly maps have less integration time at the edges, so a spatial error map is required in order to properly assess features near the noisy edges of the map.

Zeroth (M$_0$; integrated intensity), first (M$_1$; central velocity), and second (M$_2$; line-width) moment maps were made according to
\begin{equation}
M_{0} = \int{I(\nu)\ d\nu}
\end{equation}
\begin{equation}
M_{1} = \frac{1}{M_{0}}{\int{I(\nu) \nu\ d\nu}}
\end{equation}
\begin{equation}
M_{2} = \sqrt{\frac{1}{M_{0}}\int{I(\nu) (\nu - M_{1})^2\ d\nu}}
\end{equation}
where $I(\nu)$ is the intensity at a given frequency, $\nu$. The error on the zeroth moment is simply
\begin{equation}
\sigma_{M_0} = \sigma \sqrt{n}
\end{equation}
where $\sigma$ is the per-channel noise in the spectrum as calculated for our error map and $n$ is the number of spectral channels used. Errors on the first and second moment ($\sigma_{M_1}$ and $\sigma_{M_2}$) are calculated from propagation of uncertainty on the formulae for $M_{1}$ and $M_{2}$ above, but are omitted for space. 

The main choice in making moment maps lies in identifying the region of the cube to use. For the pilot survey automatic line detection was hindered by baseline ripples (particularly in worse weather) and noisy edges on the bandpasses. Improvement to the data processing pipeline are expected to mitigate baseline ripples for the full MALT90 survey and allow for automatic detection of lines, but these were not available for processing the pilot data. Therefore we use hand-identified velocities for each source to make moment maps in fixed-width windows around these velocities. Hand-identified velocities were estimated by recording the velocity at the center of each line as estimated by eye and averaging the velocities from whichever of the four main lines (\nthp, \hnc, \hcop, \hcn) were clearly detected above the noise. Where no line could be identified (53 sources), no moment map was made.

Two different velocity ranges were used for making moment maps, a narrow range for detection and a broader range for measuring line properties. A narrow velocity range ($\pm$ 2.25 km s$^{-1}$) produced the highest signal to noise measurement for weak, narrow lines by limiting the spectral region considered to the peak of the line. Typical full-width at half-maxima line-widths (\FWHM) for our sources are between 5 and 8 km s$^{-1}$ (as measured in HNC; see \S~\ref{LineProperties}), so this narrow velocity range does not adequately measure line properties. Therefore, a broader velocity range ($\pm$ 8.25 km s$^{-1}$) was also used to make moment maps. This range typically covers most of the line down to the noise, and thus comes much closer to estimating the true moments of the line. In addition, it includes the hyperfine components in both the \nthp\ and \hcn\ lines, providing a better measure of the integrated intensity for those lines; the trade-off is higher noise. We therefore report detections from the narrow velocity integration range ($\pm$ 2.25 km s$^{-1}$) and report moment information from the broader range ($\pm$ 8.25 km s$^{-1}$). 

\begin{figure}
\epsscale{1.2}
\plotone{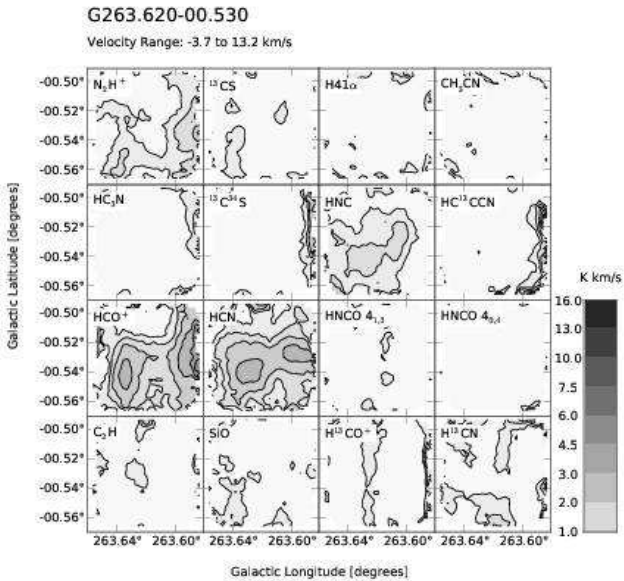}
\caption{Example integrated intensity (zeroth moment; $M_0$) map of G263.620$-$00.530. A uniform color-scale is used in all figures in this figure-set, starting at 1 K km s$^{-1}$ which is a typical 5$\sigma_{m_0}$ contour for our dataset.  Maps taken in worse weather have higher noise and the edges of the maps have higher noise, so not all emission at this level is necessarily significant. We use spatially varying noise maps to identify genuinely significant emission for the analysis presented in the text. Figure 1 of 134 in this figureset.}
\label{ExampleFigure}
\end{figure}

\begin{figure}
\epsscale{1.2}
\plotone{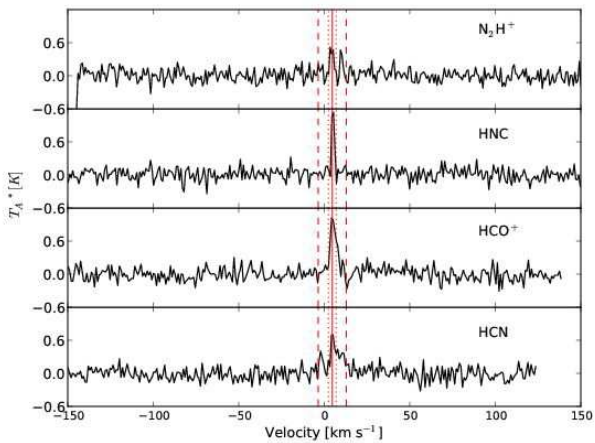}
\caption{Spectra of our four main lines at their respective positions of maximum integrated intensity (or in the center of the map for non-detections) for G263.620$-$00.530. Vertical [red] lines show the central velocity estimated by hand [solid], the small velocity range used for detection [dotted line], and larger velocity range used for measuring moments [dashed line].  Figure 1 of 182 in this figureset.}
\label{ExampleSpectra}
\end{figure}

Integrated intensity ($M_0$) maps for each source detected in any line are presented in Figure Set~\ref{ExampleFigure}. To facilitate inter-comparison, all maps are displayed on the same intensity scale, with the lowest contour at 1 K km s$^{-1}$ which is a typical 5$\sigma$ uncertainty in the integrated intensity. These moment maps are all made in a fixed velocity range around hand-identified central velocities. We show the spectra for our four main lines at their respective positions of maximum integrated intensity in Figure Set~\ref{ExampleSpectra}. For sources without any detections Figure Set~\ref{ExampleSpectra} shows the spectra at the center of the map.

Basic source properties, including our hand-determined centroid velocities and the per-channel standard deviation at the center of the map (which is representative of the fully-sampled portion of the map) are summarized in Table~\ref{basicdata}. In five cases, two distinct and widely separated velocity components were seen in a source. In these cases, each velocity component was used to create moment maps. The stronger line is listed first in Table~\ref{basicdata} and is the main line used when considering detection statistics. We did not consider these as separate sources for the detection statistics (because we are interested in knowing if a given catalog will give us a detection at a given spatial position), but did consider them as separate sources (giving us a total of 187 sources) when considering the distributions of measured moments (because they are likely two separate dense molecular clumps at distinct distances as well as velocities). 

The positions of maximum integrated intensity within each map were found by making a signal-to-noise ratio map from M$_{0}$ and $\sigma_{M_0}$, setting the poorly-sampled three edge pixels to zero, boxcar smoothing by a factor of three (i.e. taking the sliding average of three pixels), and identifying the maximum value. This process produces a potentially different maximum integrated intensity position for each line, but this is desirable as several of our sources exhibit strong spatial variation in line intensity ratios. The parameters of the four main lines (\nthp, \hnc, \hcop, \hcn) at their respective positions of maximum integrated intensity are listed in Table~\ref{mainlines}.

\section{Results}

\subsection{Detection Statistics and Line Properties}
\label{LineProperties}
The large size of our data set requires that we set a high level of significance when searching for features to avoid many false positives. With 187 sources, each with 16 lines and 31$\times$31 pixels in each map we are searching for line detections in nearly 3 million spectra. A 5$\sigma$ detection criteria should produce one false positive per 1.7 million measurements (for a perfectly normal distribution). We consider this to be an acceptably small level of contamination, and refer to a 5$\sigma$ detection as a robust detection. Additional selection criteria combined with a lower detection threshold could be used to search for additional weak lines. For instance, to improve the completeness of \htcop\ detections we could adopt a lower $\sigma$ threshold while constraining the search to locations with significant \hcop\ flux. 

\begin{figure}
\epsscale{1.3}
\plotone{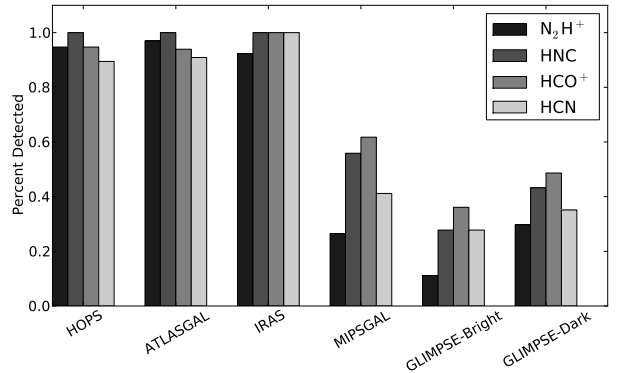}
\caption{The percentage of robust detections of the four main pilot survey lines (\nthp, \hnc, \hcop, and \hcn\ ) as a function of input catalog. A robust detection is a source with maximum integrated intensity (M$_{0}$) $> 5\sigma_{M_{0}}$, excluding the poorly sampled edge 3 pixels (27\arcsec). }
\label{detections}
\end{figure}

\begin{figure}
\epsscale{1.3}
\plotone{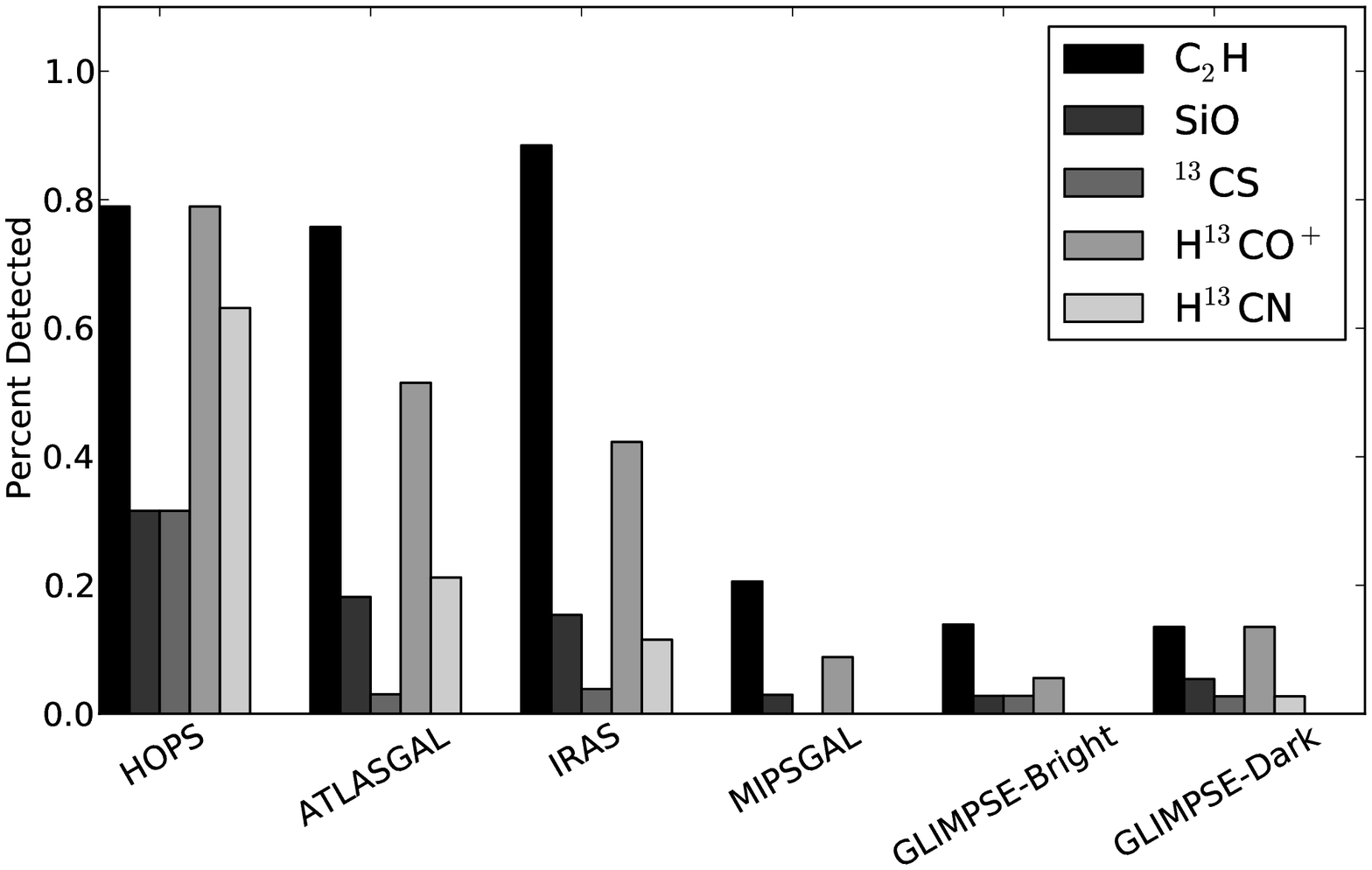}
\caption{The percentage of robust detections of the frequently detected weaker pilot survey lines (\cth, \tcs, \sio, \htcop\ and \htcn\ ) as a function of input catalog. A robust detection is a source with maximum integrated intensity (M$_{0}$) $> 5\sigma_{M_{0}}$, excluding the poorly sampled edge 3 pixels (27\arcsec). }
\label{raredetections}
\end{figure}

Our robust detection rates of the four main lines (\nthp, \hnc, \hcop, and \hcn\ ) were high ($>90\%$) for the HOPS, ATLASGAL and IRAS catalogs, and lower ($<60\%$) for sources chosen based on the morphology of \emph{Spitzer} emission (see Figure~\ref{detections}). Detection rates were comparable for the HOPS, ATLASGAL and IRAS samples, and similar for all four species. Five additional species had robust detections: \cth, \tcs, \sio, \htcop\ and \htcn. These detection rates are presented in Figure~\ref{raredetections}. \cth, in particular, was commonly seen, with detection rates between 10 and 90\% for the six different surveys. Again, the HOPS, ATLASGAL and IRAS catalogs produced more robust detections than the catalogs based on the morphology of \emph{Spitzer} emission. The low detection rates of the three input catalogs based on \emph{Spitzer} morphology are likely due to these catalogs identifying features which are not truly associated with dense clumps. For instance, 10\% to 20\% of the IRDC candidates in the \citet{Peretto:2009} catalog are not detected in the Herschel Hi-GAL survey \citep{Peretto:2010} and only 58\% of the IRDC candidates in the \citet{Simon:2006a} catalog are detected in CS \citep{Jackson:2008}. These non-detections suggest that the IRDC catalogs contain sources with a range of column densities, including sources with low column densities that do not have sufficient column density to be observed with the sensitivity limits of this survey. Table~\ref{detectiontable} presents the detections and non-detections of lines for all the sources.

\begin{figure}
\epsscale{1.25}
\plotone{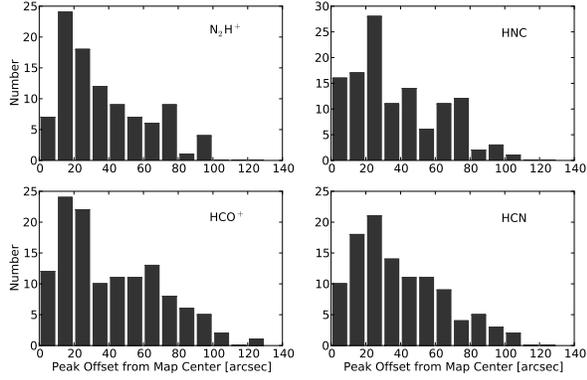}
\caption{Radial offset of maximum integrated intensity for each of the four main MALT90 pilot lines (\nthp, \hnc, \hcop, and \hcn) from the center of the map. Offsets are relative to the targeted center of the map, which is determined differently for the different input surveys. Pointing error is estimated to be less than 10\arcsec.}
\label{Positions}
\end{figure}

The detection statistics correspond to the brightest integrated emission anywhere in the map, not necessarily at the center of the map. Each input catalog provides a central position of the source, which was used as the center of the map. The positions of maximum integrated intensity tend to be clustered at the center of our images (see Figure~\ref{Positions}) with 50\% of maximum integrated intensity detections for each of the four main lines occurring within 40\arcsec\ of the map center. This suggests that the input catalog positions are good choices for the center of the map. 

\begin{figure}
\epsscale{1.3}
\plotone{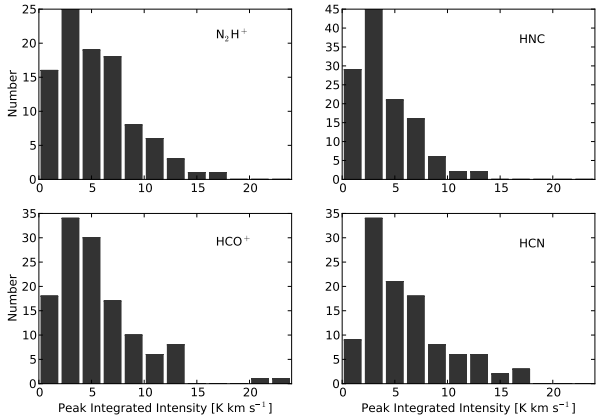}
\caption{Maximum integrated intensity histograms of the four main species. The temperature scale is T$_{A}^{*}$. Detections are incomplete below 2 K km s$^{-1}$ at the chosen 5$\sigma$ level. The distributions for the four lines are broadly similar. }
\label{mom0_hist}
\end{figure}

\begin{figure}
\epsscale{1.2}
\plotone{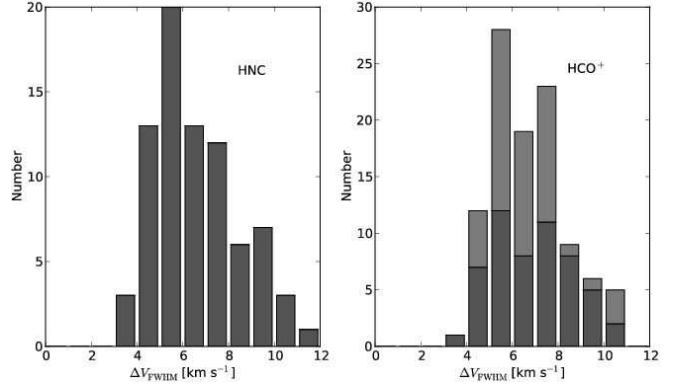}
\caption{Effective \FWHM\ of the two strong single-component lines ([left] \hnc\ and [right] \hcop) where M$_{2}$/$\sigma_{M_2} > 3$. Light gray bars on the \hcop\ histogram show points with \htcop\ detections where the \hcop\ line is likely optically thick and self-absorbed. The distributions of line-widths are similar for the two lines.}
\label{mom2_hist}
\end{figure}

The observed distributions of the integrated intensities and line-widths of the four main species (\nthp, \hnc, \hcop, and \hcn) at the brightest point in each map are displayed in Figures~\ref{mom0_hist} and \ref{mom2_hist}. Again, the detection criteria is M$_0 > 5\sigma$ in the narrow ($\pm$ 2.25 km s$^{-1}$) integration range, but the integrated intensities shown in Figures~\ref{mom0_hist} and \ref{mom2_hist} and reported in Table~\ref{mainlines} are based on the broader range ($\pm$ 8.25 km s$^{-1}$); some lines are no longer 5$\sigma$ measurements when using the broader velocity integration range. 

The integrated intensities for the four main species at their brightest location in each map show broadly similar distributions. All are incomplete below 2 K km s$^{-1}$ due to our noise and detection level. Two sources have extremely bright and broad \hcop\ lines, with M$_0 > 20$ K km s$^{-1}$, possibly indicating the presence of outflows. \hnc\ has relatively fewer lines which are both broad and bright. Although integrated intensity is a distance-dependent measurement, most sources are detected in all four lines at the same velocity (and thus distance) or in none of these lines. Therefore the similarity of integrated intensity distributions shows that the line luminosity distributions for these transitions are similar for the majority of these sources (see \S~\ref{chemicalvar} for some counter-examples). 

We calculate an effective \FWHM\ from the second moment (M$_2$) with the formula for a Gaussian profile (\FWHM\ = $\sqrt{8 \ln{2}} \times M_2$), despite the fact that many lines deviate from a Gaussian profile. \FWHM\ are often reported as a proxy for second moment, so we report this effective \FWHM\ to facilitate comparison with other studies. We report this quantity only for \hnc\ and \hcop. \nthp\ and \hcn\ are excluded because their hyperfine structure prohibits making a line-width measurement solely from our moment maps. 

We compare the \hcop\ and \hnc\ line-width distributions in Figure~\ref{mom2_hist} for sources where $M_2/\sigma_{M_2} > 3$. We break down the \hcop\ distribution based on whether \htcop\ is detected for a given source. The detection of this rare isotope typically indicates an optically thick \hcop\ line (although a non-detection of \htcop\ is not a guarantee that \hcop\ is optically thin). The distributions of \FWHM\ for \hcop\ and \hnc\ are broadly similar. We apply a two-sided Kolmogorov-Smirnov (K-S) test and find that we cannot reject the hypothesis that the \hnc\ and \hcop\ line-widths are drawn from the same population when considering just the \hcop\ line-widths in sources without a \htcop\ detection (p-value = 9\%) or all the \hcop\ line-widths (p-value = 32\%).

\begin{figure}
\epsscale{1.3}
\plotone{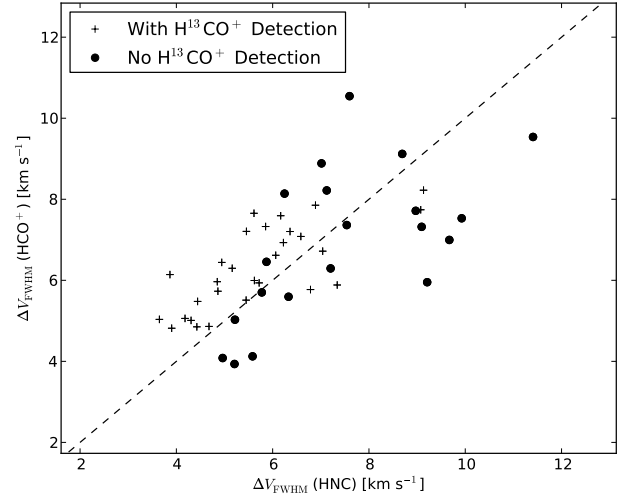}
\caption{Comparison of the effective \FWHM\ of the two strong single-component lines ([left] \hnc\ and [right] \hcop) where the positions of maximum integrated intensity are within 13\arcsec\ and M$_{2}$/$\sigma_{M_2} > 3$. The dashed line is unity. Crosses indicate sources with \htcop\ detections where the \hcop\ line is likely optically thick and self-absorbed; in these sources the \hcop\ linewidth tends to have larger than the \hnc\ linewidth. For the sources without \htcop\ detections, the linewidths are on average the same for both \hcop\ and \hnc. }
\label{mom2_comp}
\end{figure}

The line-width distributions shows in Figure~\ref{mom2_hist} are the line-widths at the positions of maximum integrated intensity for each molecular line transition. If we restrict our analysis to sources for which the positions of maximum integrated intensity for \hcop\ and \hnc\ are within one 9\arcsec\ pixel (within 13\arcsec\ to include diagonally adjacent pixels) we can compare line-widths at roughly the same position. Figure~\ref{mom2_comp} shows the results of this comparison for sources where $M_2/\sigma_{M_2} > 3$ in both lines. For the sources with an \htcop\ detection (i.e. where \hcop\ is likely to be optically thick) the \hcop\ line-width is typically larger than the \hnc\ line-width ($\langle \FWHM(\hcop)- \FWHM(\hnc) \rangle = 1.15$). For the sources without an \htcop\ detection, the line-width ratios are correlated and centered around unity ($\langle \FWHM(\hcop)- \FWHM(\hnc) \rangle = 0.98$). This suggests that the sources with \htcop\ detections do have optically thick \hcop\ emission, and that this is what produces their larger line-widths. In the pilot survey, we have no rare isotopologue of \hnc\ to study where \hnc\ might be optically thick, but the main MALT90 survey will include HN$^{13}$C instead of \htcn\ (\hcn, because of its hyperfine structure, is less likely to be optically thick for similar line intensities).

\subsection{Choice of Input Catalog}
\label{Choice}

The first goal of the MALT90 pilot survey was to select an input catalog for the full MALT90 survey. Of the 6 catalogs tested, only HOPS, ATLASGAL and IRAS had sufficiently high ($>90\%$) detection rates to be used as input catalogs for the main MALT90 survey. There are significant scientific and logistical benefits to using a single catalog when selecting sources. The main advantages are the ability to use a simple and uniform criteria for choosing sources, the ability to compare our observed line properties against properties of the input catalog, and the ability to describe the significance of non-detections. The ATLASGAL catalog provides the optimal source list for MALT90. There are three major factors in favor of using ATLASGAL: (1) catalog size, (2) a broad range of source positions and velocities, and (3) a range of evolutionary states. 

ATLASGAL provides a much larger catalog than HOPS or the \citet{Beltran:2006} survey of IRAS sources. \citet{Schuller:2009} report from the initial results of the survey about 6000 sources brighter than 0.25 Jy in 95 deg$^2$ in the Galactic range $-30^\circ \leq l \leq +11.5^\circ$ and $+15^\circ \leq l \leq +21^\circ$ with $|b| \leq 1^\circ$. In contrast the \citet{Beltran:2006} survey contains 235 sources and HOPS \citep{Walsh:2008} is expected to contain a few hundred bright \amm\ sources. Neither HOPS nor IRAS contains a sufficient number of sources for the science goals of MALT90.

\begin{figure*}
\epsscale{1.15}
\plotone{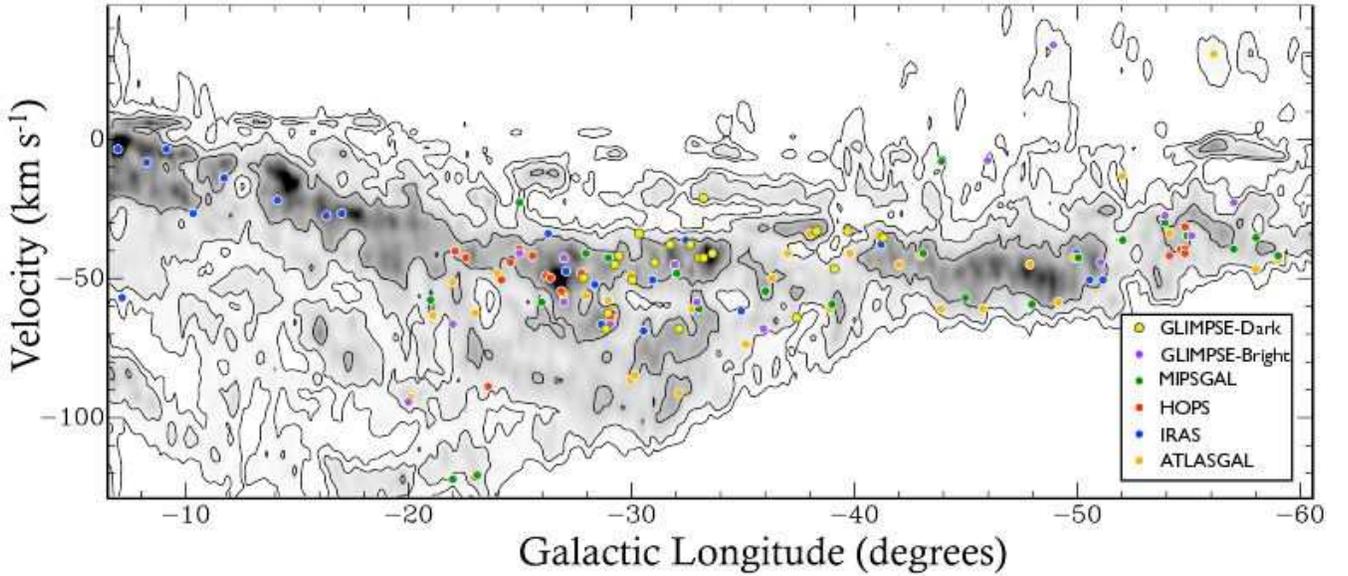}
\caption{Distribution of MALT90 pilot sources in velocity and Galactic longitude, overlaid on the Galactic CO distribution from \citet{Dame:2001} integrated over Galactic latitude. The CO contours are at 0.3, 1, 3, and 10 K arcdeg.  Different input catalogs are labelled as follows: [yellow] = Dark GLIMPSE source, [purple] = Bright GLIMPSE source, [green] = MIPSGAL source, [red] = HOPS source, [blue] = IRAS source, [orange] = ALTASGAL source. Sources with no detected line emission (and thus no velocity) are omitted. In addition, four IRAS sources with Galactic longitude between -70 and -100 degrees were omitted to display the remaining sources at a larger scale. These four omitted sources all also lie within the CO 0.3 K arcdeg emission contour.}
\label{lvdistribution}
\end{figure*}

ATLASGAL sources appear to sample many Galactic structures. Figure~\ref{lvdistribution} shows the Galactic longitude and velocity of sources with detections (using hand-determined velocities) plotted on the \citet{Dame:2001} CO map. The \citet{Dame:2001} CO map is presented as a longitude-velocity diagram integrated over $-2^\circ < b < 2^\circ$ and has units of K arcdeg. The positions of our sources in this plot all fall within the 0.3 K arcdeg CO contour and most cluster in the portions of stronger CO emission, as expected for dense, star-forming gas. The presence of ATLASGAL sources at many positions in the longitude-velocity diagram indicates that this catalog is detecting sources in a range of Galactic locations.

ATLASGAL sources cover a range of evolutionary states. The initial results of ATLASGAL found two-thirds of the sources do not have a mid- or far-infrared counterpart in the Midcourse Space Experiment (MSX) and IRAS catalogs. A closer inspection of IRAS/MSX dark sources in more sensitive \emph{Spitzer} GLIMPSE/MIPSGAL images reveals many of them associated with weaker infrared sources but still a considerable fraction of the submillimeter emission appear dark in the \emph{Spitzer} images \citep[e.g., Fig. 13 and 14 of][]{Schuller:2009} and work on the first compact source release catalog demonstrates that ATLASGAL will provide enough sources in each evolutionary state as assessed by the \emph{Spitzer} emission morphology scheme shown in Figure~\ref{spitzer}.

We thus choose ATLASGAL as the sole input catalog for the MALT90 survey because it meets all our requirements for a source list. ATLASGAL sources had high detection rates in this pilot survey, include a diversity of evolutionary states, and cover a broad range of Galactic positions. Choosing ATLASGAL as the sole input catalog also provides the benefits of having a single uniform catalog when selecting sources. We can choose our sources with a single uniform criteria and compare our MALT90 measurements against ATLASGAL catalog properties such as the flux and extent of 870~\micron\ emission. 

\subsection{MALT90 Survey Strategy}

The second goal MALT90 pilot survey was to test the observing set up and verify that it allows us to achieve our science goals. MALT90 is fundamentally a mapping survey; although some science goals (such as determining distances to clumps) could be achieved with a single pointing, the majority of our science goals rely on maps. Our configuration allows us to map sources in multiple lines at high spectral resolution. Maps of multiple lines allows us to study the chemical variation within a clump, which is most useful if clumps are typically spatially resolved and at least sometimes exhibit strong chemical variation. Mapping at high velocity resolution (0.11 km/s) allows us to study spatial variation in line profiles which may indicate changes in the strength of turbulence, large scale motions (rotation, shear, infall or outflow), or multiple velocity components. The pilot survey allowed us to verify that our on-the-fly maps had sufficient sensitivity and that we could make maps without significant artifacts. 

\subsubsection{Mapping Multiple Dense-gas Tracers to Reveal Chemistry}
\label{chemicalvar}
\begin{figure*}
\plottwo{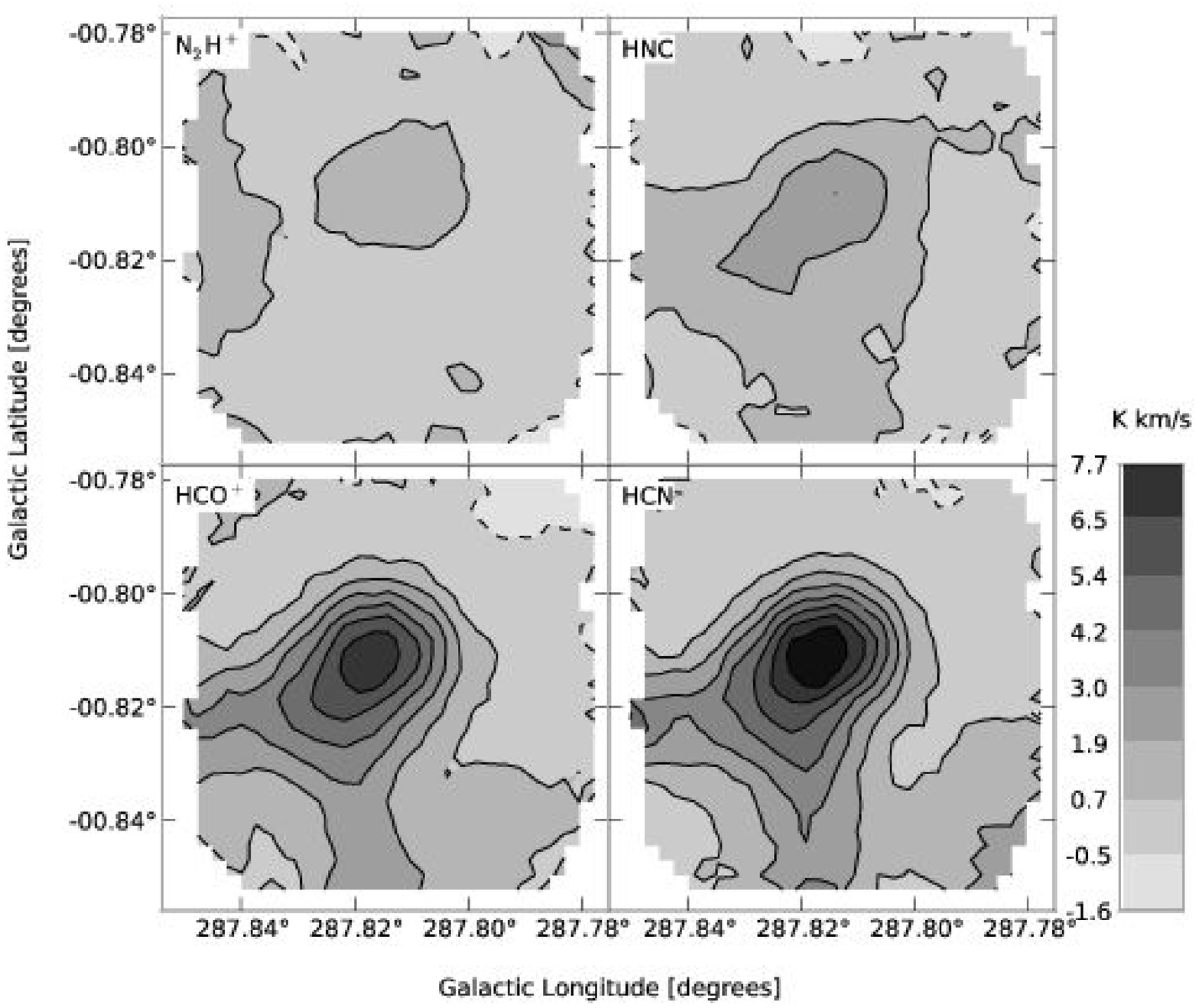}{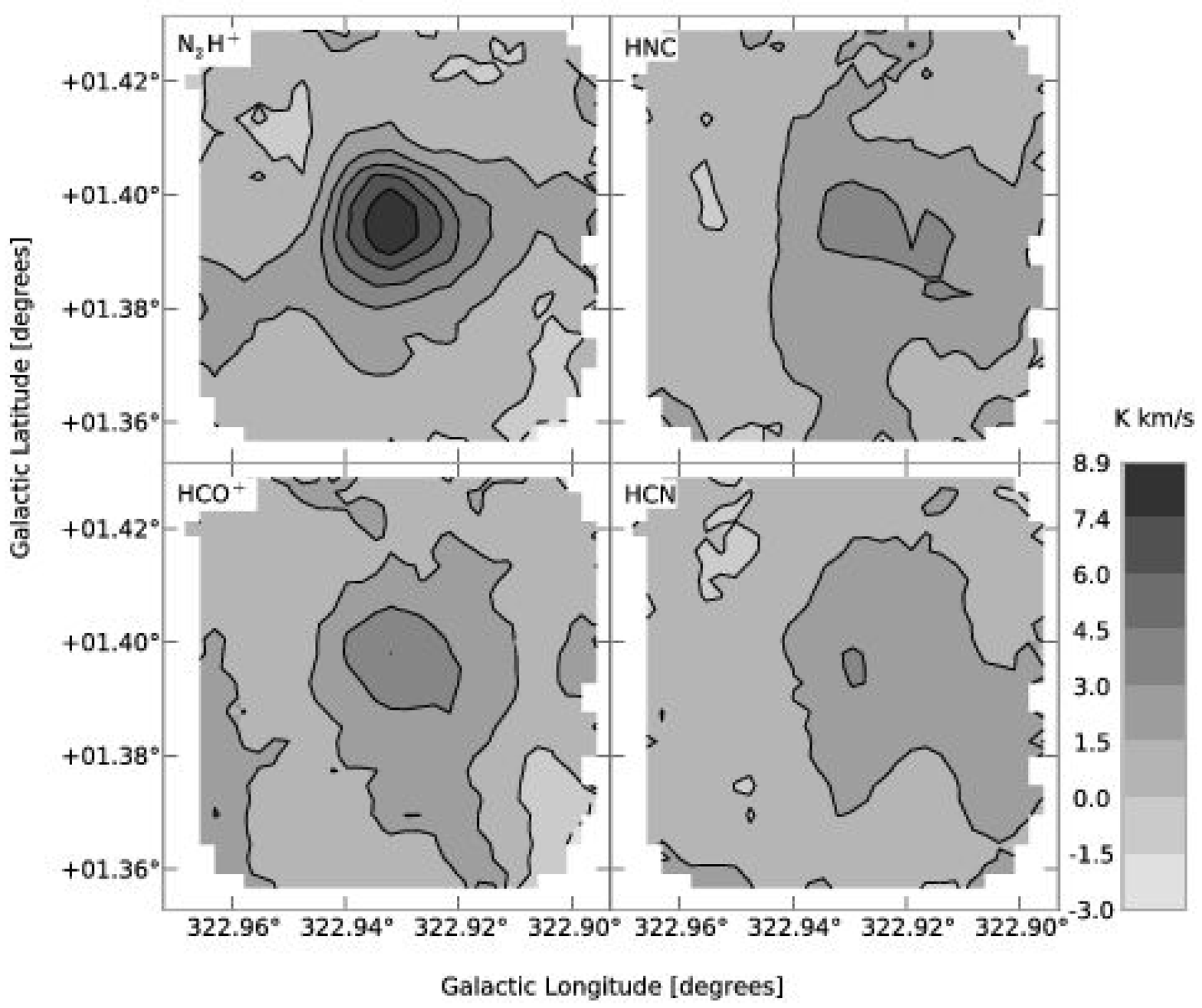}
\caption{Two examples of strong chemical variation in \nthp. Left: G287.814$-$00.816 shows strong (3 - 9 K km s$^{-1}$) \hnc, \hcop, and \hcn\ lines, but weak (1.65$\pm$0.25 km s$^{-1}$) \nthp. Right: G322.932$+$01.393 shows a very strong \nthp\ line (8.93$\pm$ 0.33 K km s$^{-1}$), but comparatively weak \hnc\ and \hcop\ (3.5 - 4.5 K km s$^{-1}$) and a non-detection of \hcn ($< 1.66$ K km s$^{-1}$).}
\label{nthp_odd}
\end{figure*}

We see strong chemical variation in the MALT90 pilot sources which validates our decision to map multiple lines in these sources. All four of our main lines (\nthp\,\hcn\, \hcop\ and \hnc) are ground state transitions of molecules with similar critical densities. As tracers of dense gas, the emission from these lines typically show similar morphologies in MALT90 pilot sources, but this is not always the case. Figure~\ref{nthp_odd} shows two example sources where \nthp\ varies significantly with respect to the other species. Figure~\ref{nthp_odd} shows one source where the maximum \nthp\ integrated intensity is a factor of 4-8 times weaker than the other three main lines. Conversely, the other source in Figure~\ref{nthp_odd} has \nthp\ emission which is twice as strong as that of \hcn, \hcn\ or \hcop. Large variation in the \hcop/\nthp\ integrated intensity ratio in high-mass star-forming regions has been noted before \citep[e.g.,][]{Turner:1977,Walsh:2006}. Thus, the combination of mapping in several lines simultaneously gives us the most complete picture of the spatial distribution of the various molecules, which is crucial to studying variations in the chemistry with the MALT90 sources.

\subsubsection{Mapping Strategy}

\begin{figure}
\epsscale{1.2}

\plotone{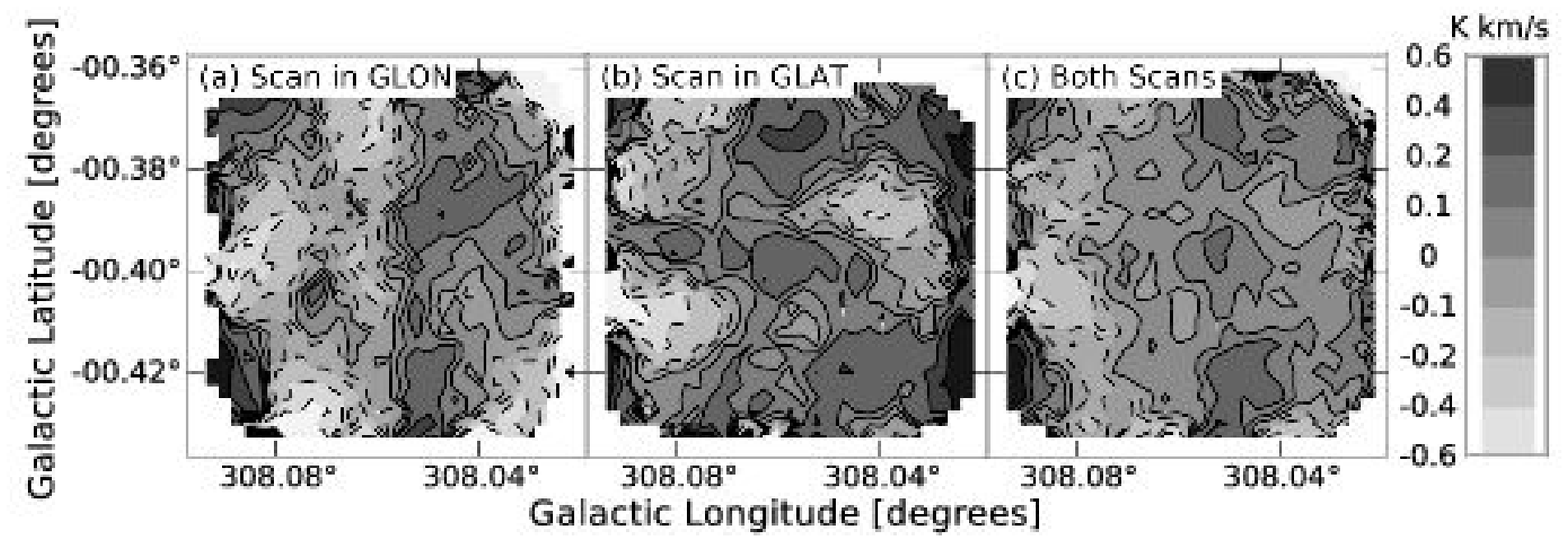}
\caption{Zeroth moment (integrated intensity) maps of \tcts\ in G308.058$-$00.397. \tcts\ has very low abundance, so these maps show only noise. In panel (a) the map was made with scans of constant Galactic longitude, which is the default mode for the MALT90 pilot survey. In panel (b) the map was made with scans of constant Galactic latitude. In panel (c), the two individual maps were combined, reducing the striping artifacts visible in maps (a) and (b). This reduction of artificial structure motivates combining maps scanned in both directions in the full MALT90 Survey. }
\label{CrossMap}
\end{figure}

Mapping artifacts are seen in many of our maps, typically manifesting as stripes in the direction of scans. Our sources were generally mapped with scans of constant Galactic longitude so that each strip in Galactic longitude uses the same reference spectrum. Noise or gain variations in this reference spectrum can therefore produce stripes in the map, and this phenomenon is particularly prevalent in sources observed in bad weather. We chose to map using scans of constant Galactic longitude for the pilot survey because most extended structures in the Galactic plane are parallel to the plane. Thus, noise stripes are easier to identify, since they typically run perpendicular to real features. We mapped two sources (G305.887$+$00.016 and G308.058$-$00.397) in both Galactic longitude and latitude to see if this would mitigate the striping. Figure~\ref{CrossMap} shows the two individual maps for G308.058$-$00.397 as well as the combination of both maps for a signal-free \tcts\ cube. The striping visible in the scan direction in both individual maps is significantly reduced in the combined image. We therefore decided to map using scans of both constant Galactic longitude and latitude in the full MALT90 survey.

\subsubsection{The Value of High-resolution Velocity Information}

\begin{figure*}
\epsscale{1.15}

\plotone{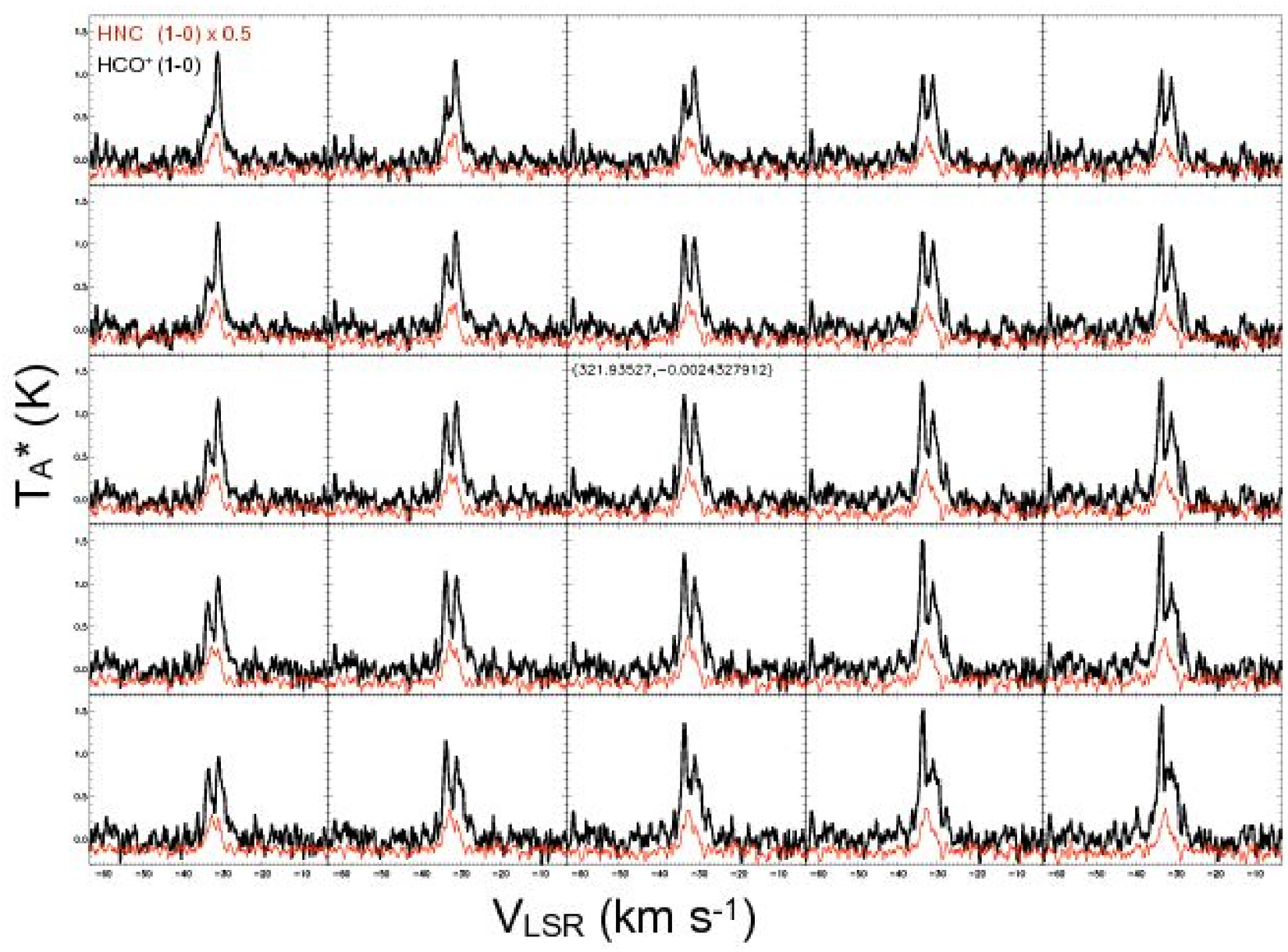}
\caption{\hcop\ [thick/black] and \hnc\ [thin/red] line profiles over the central portion of the G321.935$-$00.007 cube. The \hnc\ spectra has been scaled by a factor of 0.5.The \hcop\ line shows a strong dip at the systemic velocity, but the relative strengths of the blue and red wings vary throughout this map. Adjacent spectra are separated by 9\arcsec.}
\label{Infall}
\end{figure*}

Figure~\ref{Infall} shows the central portion of the maps for the source G321.935$-$00.007 in \hnc\ and \hcop. The \hcop\ line shows self-absorption at the systemic velocity (traced by the \hnc\ which shows little or no non-gaussianity). In the lower-right portion of the map, this self-absorption shows an asymmetric profile with brighter blue-shifted emission. Such a profile is characteristic of infall of cold gas toward a hot central source \citep[e.g.,][]{Mardones:1997}. This characteristic shape is not present in the upper-left of the map, where we see a red-shifted profile usually associated with expansion. It is possible that infall is happening only in part of this source or that other kinematic complexity is present; the large variance of this complex line shape over the source demonstrates the value of mapping at high velocity resolution.

\subsubsection{Studying Different Stages of Evolution}
\label{evolution}
The many lines observed in MALT90 provide information about the evolutionary state of the clumps observed. This information can be combined with \emph{Spitzer} morphological classification and dust temperature determination from spectral-energy distribution fitting \citep[e.g.,][]{Rathborne:2010, Peretto:2010} to constrain the evolutionary state of a clump. As a short example we present three sources in different \emph{Spitzer} morphological states and show what information can be gained from the molecular lines in each case. 

\begin{figure}
\epsscale{1.15}
\plotone{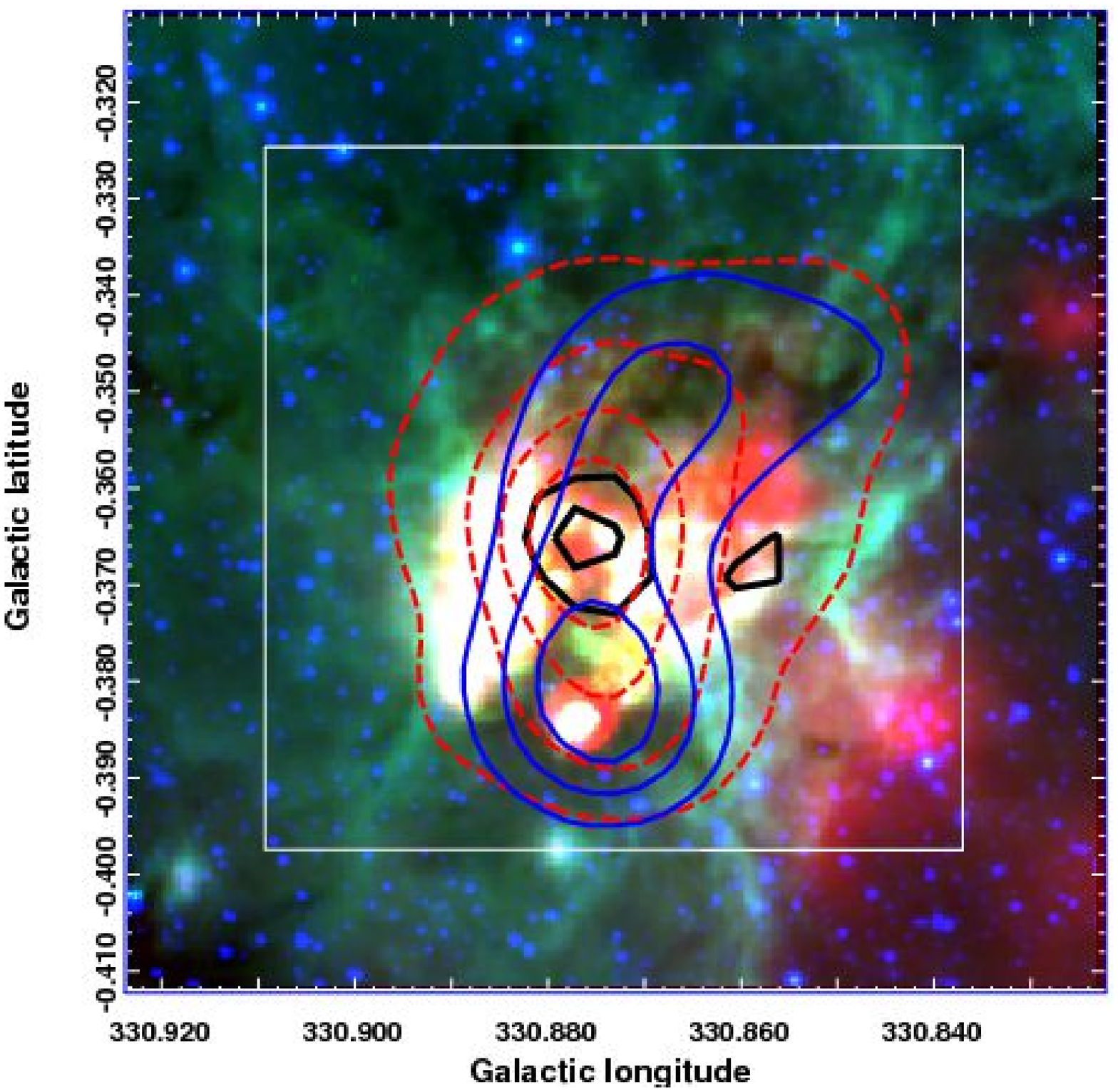}
\caption{An example \HII\ clump (G330.873$-$00.361) shown as a three color image with \emph{Spitzer}/IRAC band 1 (3.6~\micron) and band 4 (8.0~\micron) from GLIMPSE in blue and green and \emph{Spitzer}/MIPS band 1 (24~\micron) from MIPSGAL in red. The white box shows the extent of the MALT90 pilot survey map. Dashed (red) contours are \hcop\ integrated intensity (plotted at SNR of 15, 30, 45, 60; $\sigma \sim$ 0.25 K km s$^{-1}$). Thin (blue) contours are \nthp\ integrated intensity (plotted at SNR of 10, 20, 30, 40; $\sigma \sim$ 0.26 K km s$^{-1}$). Thick (black) contours are \chtcn\ integrated intensity (plotted at SNR of 2, 4; $\sigma \sim$ 0.28 K km s$^{-1}$). The presence of \chtcn\ identifies the hot core associated with a massive protostar. This location is also the position of maximum integrated intensity for most of the other molecules (\hcn, \hnc, \sio, \tcs), but \nthp\ peaks in the south as the position of another 24~\micron\ point source. }
\label{ExampleHII}
\end{figure}

Figure~\ref{ExampleHII} shows G330.873$-$00.361, a clump drawn from the HOPS survey catalog and classified as an \HII\ region from the \emph{Spitzer} GLIMPSE/MIPSGAL images due to its strong extended 8 and 24~\micron\ emission. The presence of \chtcn\ ($J_K = 5_1 - 4_1$) emission identifies the hot core associated with a massive protostar since this molecule is seen only in warm ($T >$ 100 K) and dense ($n >$ 10$^{5}$ cm$^{-3}$) regions and is often detected in \HII\ regions \citep[e.g.,][]{Purcell:2006}. The location of the core is also the position of maximum integrated intensity for most of the other molecules (\hcn, \hnc, \sio, \tcs), but not \nthp, which peaks in the south at the position of a 24~\micron\ point source. Although this source is clearly identified as an \HII\ region from the characteristic \emph{Spitzer} appearance of a 24~\micron\ inner bubble surrounded by a ring of 8~\micron\ emission \citep[e.g.,][]{Watson:2008}, the detection of a hot core molecule identifies the location of the central exciting source. The chemical difference (\nthp/\hcop\ ratio) between the southern source suggests either a different luminosity for the exciting source or a different evolutionary state. 

\begin{figure}
\epsscale{1.15}
\plotone{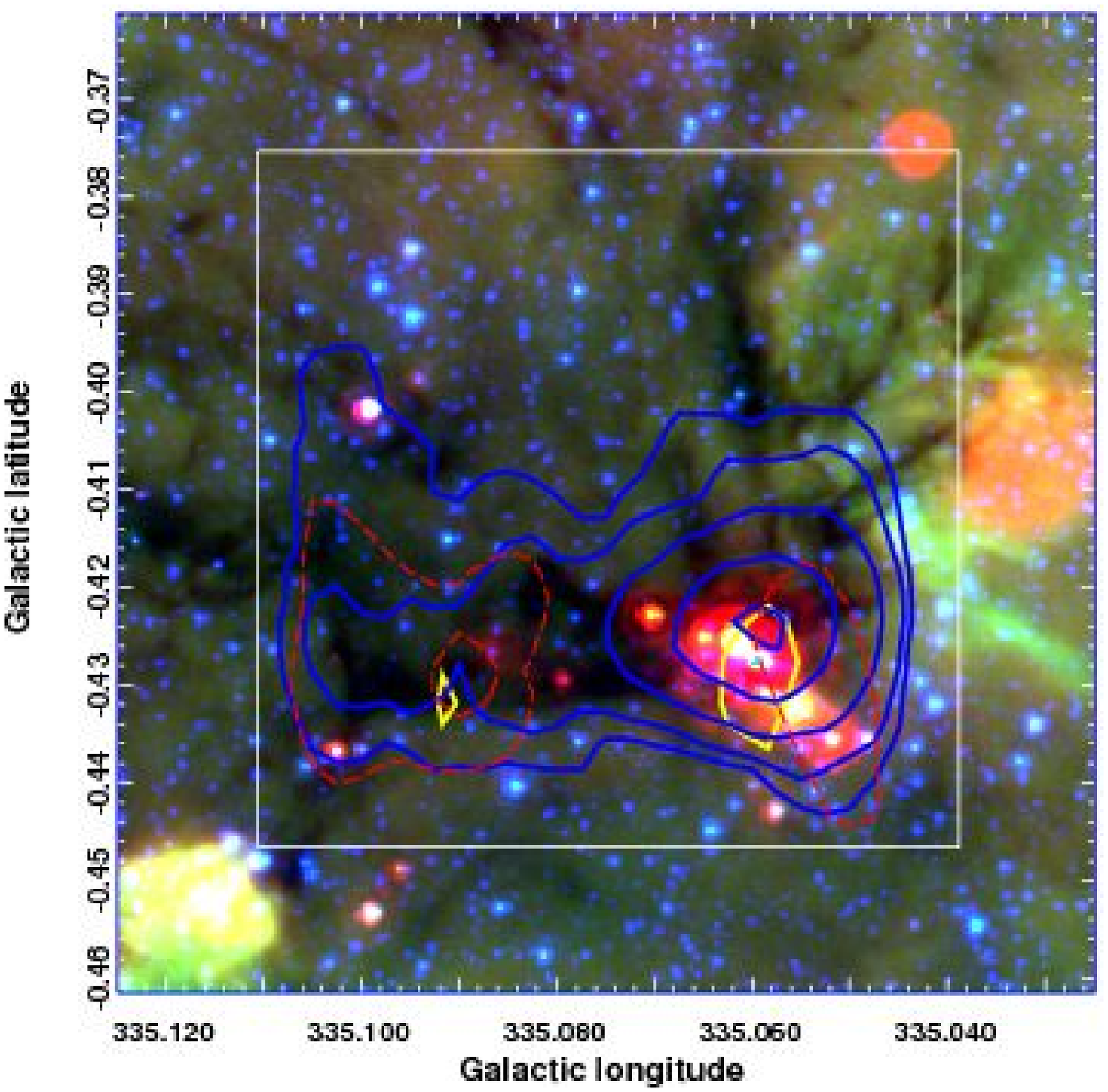}
\caption{An example protostellar clump (G335.075$-$00.411) shown as a three color image with \emph{Spitzer}/IRAC band 1 (3.6~\micron) and band 4 (8.0~\micron) from GLIMPSE in blue and green and \emph{Spitzer}/MIPS band 1 (24~\micron) from MIPSGAL in red. The white box shows the extent of the MALT90 pilot survey map. Dashed (red) contours are \hcop\ integrated intensity (plotted at SNR of 5, 7; $\sigma \sim$ 0.30 K km s$^{-1}$). Thin (blue) contours are \nthp\ integrated intensity (plotted at SNR of 7, 9, 13, 19, 25; $\sigma \sim$ 0.29 K km s$^{-1}$). Thick (yellow) contours are \sio\ integrated intensity (plotted at SNR of 4; $\sigma \sim$ 0.20 K km s$^{-1}$ in the small velocity window). This clump was drawn from the HOPS catalog and classified as protostellar due to the presence of several 24~\micron\ point sources. The strongest 24~\micron\ point sources are associated with \sio\ emission at the same velocity as the clump, indicating the presence of protostellar outflows in the clump.}
\label{ExampleProtostellar2}
\end{figure}

Figure~\ref{ExampleProtostellar2} shows G335.075$-$00.411, a clump drawn from the HOPS survey catalog and classified as protostellar from the \emph{Spitzer} GLIMPSE/MIPSGAL images due to the presence of 24~\micron\ point sources within a dark extinction feature without extended 8 or 24~\micron\ emission. The spatial coincidence of the 24~\micron\ point sources and the 8~\micron\ extinction feature suggests that the 24~\micron\ point sources are associated with the clump, and the MALT90 pilot survey data confirms this association. The \nthp\ integrated intensity contours trace the 8~\micron\ emission extinction feature. There is \sio\ emission at the same velocity at the position of the brightest 24~\micron\ point source. Since \sio\ emission is normally associated with outflow activity in protostars \citep[e.g.,][]{Lopez-Sepulcre:2011}, a detection of this line at the same velocity as the clump is strong evidence that the 24~\micron\ point source is associated with this clump. 

\begin{figure*}
\epsscale{1.1}
\plotone{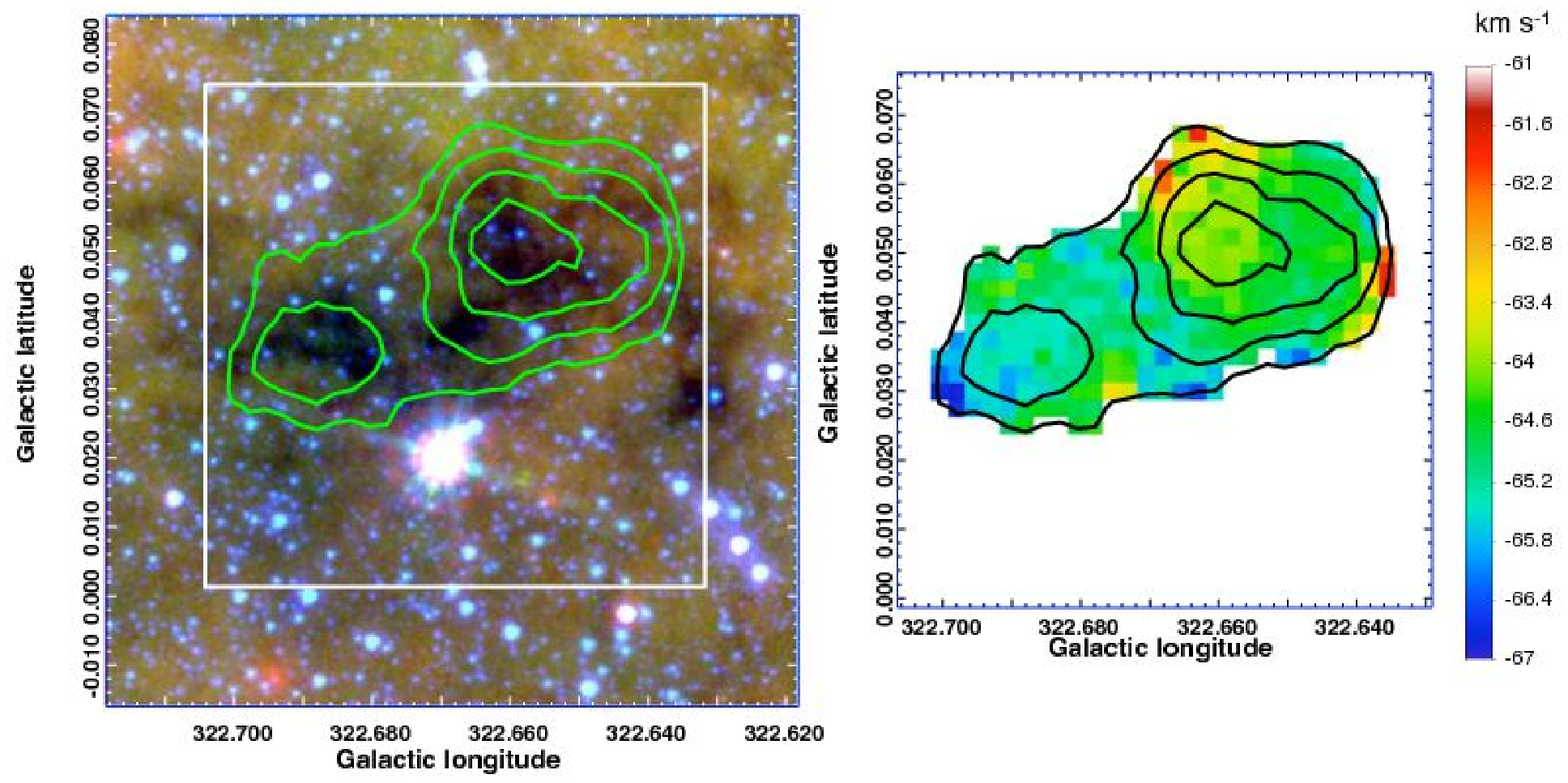}
\caption{An example quiescent clump (G322.668$-$00.038) shown (left) as a three color image with \emph{Spitzer}/IRAC band 1 (3.6~\micron) and band 4 (8.0~\micron) from GLIMPSE in blue and green and \emph{Spitzer}/MIPS band 1 (24~\micron) from MIPSGAL in red. The white box shows the extent of the MALT90 pilot survey map. The extinction seen at 8.0~\micron\ is well traced by the HNC integrated intensity contours (plotted as SNR of 4, 6, 8, 10; $\sigma \sim$ 0.25 K km s$^{-1}$) which show two distinct peaks. The right panel shows the centroid velocity M$_1$ in color with the HNC integrated intensity contours. The two peaks are at very similar velocities (-64 km s$^{-1}$ and -65.5 km s$^{-1}$, suggesting that the extinction feature is all at the same distance.}
\label{ExampleQuiescent}
\end{figure*}

Figure~\ref{ExampleQuiescent} shows G322.668$-$00.038, a clump drawn from the GLIMPSE-Dark catalog and classified as quiescent from the \emph{Spitzer} GLIMPSE/MIPSGAL images due to the lack of 8 or 24~\micron\ emission inside the 8~\micron\ extinction feature. As a quiescent clump in the early stages of evolution, this object displays less complex chemistry than clumps in more evolved stages with only the four main lines (\nthp, \hnc, \hcop, and \hcn) detected. The \hnc\ integrated intensity emission shows two distinct peaks associated with two of the darkest 8~\micron\ extinction features. The velocity field of \hnc\ shows that these two peaks are at very similar velocities (-64 km s$^{-1}$ and -65.6 km s$^{-1}$), strongly suggesting that both peaks are at the same distance and that the entire extinction feature is a single physical object. We use the \citet{Clemens:1985} rotation curve to calculate a kinematic distance for this clump; the near distance is 4.27 kpc and the far distance is 9.25 kpc. Because we see the clump as an extinction feature against the diffuse Galactic background, it is reasonable to assume that the near distance is correct. The MALT90 map therefore allows us (1) to identify which extinction features are likely a single physical object versus a chance projection and (2) to assign a distance which is useful for any further study of this object. 

\section{Conclusion}

We have described the MALT90 pilot survey, carried out to demonstrate the feasibility of the MALT90 survey, identify the best input catalog for choosing MALT90 targets, and optimize the survey parameters. We choose the ATLASGAL \citep{Schuller:2009} catalog as our source list on the basis of its high detection rates for the main four survey lines ($>$ 90\% for \nthp, \hnc, \hcop, and \hcn) and the large number of dense molecular clumps in different evolutionary stages in this catalog. The surveys which provided a prior selection for regions of high column density, either from optically-thin dust (ATLASGAL at 870~\micron, the \citet{Beltran:2006} survey at 1.2-mm) or another dense gas tracer (NH$_3$ from HOPS) produced much higher detection rates than choosing sources identified based on \emph{Spitzer} emission morphology without this prior.

We have briefly summarized the data obtained from the MALT90 pilot survey and highlighted some of the science possible with this survey including studying chemical variation, the kinematics of massive dense clumps and the galactic distribution of dense molecular clumps associated with high-mass star formation. We have made the full data-set publicly available through this publication and the MALT90 website, including reduced data-cubes and uniform moment maps which facilitate easy inspection of the data. This collection is already one of the largest sets of 90 GHz molecular line maps for dense molecular clumps. The full survey is underway and plans to map a total of 3,000 candidate dense molecular clumps, increasing this sample by an order of magnitude and providing a valuable database for studying many aspects of high-mass star formation.

\section{Acknowledgements}
The Mopra telescope is part of the Australia Telescope and is funded by the Commonwealth of Australia for operation as National Facility managed by CSIRO. The UNSW-MOPS Digital Filter Bank used for the observations with the Mopra telescope was provided with support from the Australian Research Council, together with the University of New South Wales, University of Sydney and Monash University. This research has made use of the NASA/ IPAC Infrared Science Archive (for access to GLIMPSE and MIPSGAL images), which is operated by the Jet Propulsion Laboratory, California Institute of Technology, under contract with the National Aeronautics and Space Administration. This research has made use of NASA's Astrophysics Data System Bibliographic Services. JMJ gratefully acknowledges funding support from NSF Grant No. AST-0808001. The MALT90 project team gratefully acknowledges the use of dense core positions supplied by ATLASGAL. ATLASGAL is a collaboration between the Max Planck Gesellschaft (MPG: Max Planck Institute for Radioastronomy, Bonn and the Max Planck Institute for Astronomy, Heidelberg), the European Southern Observatory (ESO) and the University of Chile. Thanks to Anita Titmarsh and the duty astronomers and staff at the Paul Wild Observatory for their assistance during the observations.

\clearpage

\bibliography{}

\begin{thebibliography}{36}

\expandafter\ifx\csname natexlab\endcsname\relax\\def\natexlab#1{#1}\fi
        \expandafter\ifx\csname bibnamefont\endcsname\relax
          \def\bibnamefont#1{#1}\fi
        \expandafter\ifx\csname bibfnamefont\endcsname\relax
          \def\bibfnamefont#1{#1}\fi
        \expandafter\ifx\csname citenamefont\endcsname\relax
          \def\citenamefont#1{#1}\fi
        \expandafter\ifx\csname url\endcsname\relax
          \def\url#1{\texttt{#1}}\fi
        \expandafter\ifx\csname urlprefix\endcsname\relax\def\urlprefix{URL }\fi
        \providecommand{\bibinfo}[2]{#2}
        \providecommand{\eprint}[2][]{\url{#2}}
        
\bibitem[{{Beltr{\'a}n} {et~al.}(2006){Beltr{\'a}n}, {Brand}, {Cesaroni},  {Fontani}, {Pezzuto}, {Testi}, \& {Molinari}}]{Beltran:2006}
{Beltr{\'a}n}, M.~T., {Brand}, J., {Cesaroni}, R., {et~al.} 2006, \aap, 447, 221

\bibitem[{{Benjamin} {et~al.}(2003){Benjamin}, {Churchwell}, {Babler}, {Bania},  {Clemens}, {Cohen}, {Dickey}, {Indebetouw}, {Jackson}, {Kobulnicky},  {Lazarian}, {Marston}, {Mathis}, {Meade}, {Seager}, {Stolovy}, {Watson},  {Whitney}, {Wolff}, \& {Wolfire}}]{Benjamin:2003}
{Benjamin}, R.~A., {Churchwell}, E., {Babler}, B.~L., {et~al.} 2003, \pasp, 115, 953

\bibitem[{{Bergin} {et~al.}(2001){Bergin}, {Ciardi}, {Lada}, {Alves}, \&  {Lada}}]{Bergin:2001}
{Bergin}, E.~A., {Ciardi}, D.~R., {Lada}, C.~J., {Alves}, J., \& {Lada}, E.~A.  2001, \apj, 557, 209

\bibitem[{{Bergin} \& {Tafalla}(2007)}]{Bergin:2007}
{Bergin}, E.~A. \& {Tafalla}, M. 2007, \araa, 45, 339

\bibitem[{{Brown} {et~al.}(1988){Brown}, {Charnley}, \& {Millar}}]{Brown:1988}
{Brown}, P.~D., {Charnley}, S.~B., \& {Millar}, T.~J. 1988, \mnras, 231, 409

\bibitem[{{Carey} {et~al.}(2009){Carey}, {Noriega-Crespo}, {Mizuno}, {Shenoy},  {Paladini}, {Kraemer}, {Price}, {Flagey}, {Ryan}, {Ingalls}, {Kuchar},  {Pinheiro Gon{\c c}alves}, {Indebetouw}, {Billot}, {Marleau}, {Padgett},  {Rebull}, {Bressert}, {Ali}, {Molinari}, {Martin}, {Berriman}, {Boulanger},  {Latter}, {Miville-Deschenes}, {Shipman}, \& {Testi}}]{Carey:2009}
{Carey}, S.~J., {Noriega-Crespo}, A., {Mizuno}, D.~R., {et~al.} 2009, \pasp, 121, 76

\bibitem[{{Clemens}(1985)}]{Clemens:1985}
{Clemens}, D.~P. 1985, \apj, 295, 422

\bibitem[{{Dame} {et~al.}(2001){Dame}, {Hartmann}, \& {Thaddeus}}]{Dame:2001}
{Dame}, T.~M., {Hartmann}, D., \& {Thaddeus}, P. 2001, \apj, 547, 792

\bibitem[{{Fazio} {et~al.}(2004){Fazio}, {Hora}, {Allen}, {Ashby}, {Barmby},  {Deutsch}, {Huang}, {Kleiner}, {Marengo}, {Megeath}, {Melnick}, {Pahre},  {Patten}, {Polizotti}, {Smith}, {Taylor}, {Wang}, {Willner}, {Hoffmann},  {Pipher}, {Forrest}, {McMurty}, {McCreight}, {McKelvey}, {McMurray}, {Koch},  {Moseley}, {Arendt}, {Mentzell}, {Marx}, {Losch}, {Mayman}, {Eichhorn},  {Krebs}, {Jhabvala}, {Gezari}, {Fixsen}, {Flores}, {Shakoorzadeh}, {Jungo},  {Hakun}, {Workman}, {Karpati}, {Kichak}, {Whitley}, {Mann}, {Tollestrup},  {Eisenhardt}, {Stern}, {Gorjian}, {Bhattacharya}, {Carey}, {Nelson},  {Glaccum}, {Lacy}, {Lowrance}, {Laine}, {Reach}, {Stauffer}, {Surace},  {Wilson}, {Wright}, {Hoffman}, {Domingo}, \& {Cohen}}]{Fazio:2004}
{Fazio}, G.~G., {Hora}, J.~L., {Allen}, L.~E., {et~al.} 2004, \apjs, 154, 10

\bibitem[{{Fuller} {et~al.}(2005){Fuller}, {Williams}, \&  {Sridharan}}]{Fuller:2005}
{Fuller}, G.~A., {Williams}, S.~J., \& {Sridharan}, T.~K. 2005, \aap, 442, 949

\bibitem[{{Gerin} {et~al.}(2011){Gerin}, {Ka{\'z}mierczak}, {Jastrzebska},  {Falgarone}, {Hily-Blant}, {Godard}, \& {de Luca}}]{Gerin:2011}
{Gerin}, M., {Ka{\'z}mierczak}, M., {Jastrzebska}, M., {et~al.} 2011, \aap, 525, A116+

\bibitem[{{Gibson} {et~al.}(2009){Gibson}, {Plume}, {Bergin}, {Ragan}, \&  {Evans}}]{Gibson:2009}
{Gibson}, D., {Plume}, R., {Bergin}, E., {Ragan}, S., \& {Evans}, N. 2009,  \apj, 705, 123

\bibitem[{{Hirota} {et~al.}(1998){Hirota}, {Yamamoto}, {Mikami}, \&  {Ohishi}}]{Hirota:1998}
{Hirota}, T., {Yamamoto}, S., {Mikami}, H., \& {Ohishi}, M. 1998, \apj, 503,  717

\bibitem[{{Jackson} {et~al.}(2008){Jackson}, {Finn}, {Rathborne}, {Chambers},  \& {Simon}}]{Jackson:2008}
{Jackson}, J.~M., {Finn}, S.~C., {Rathborne}, J.~M., {Chambers}, E.~T., \&  {Simon}, R. 2008, \apj, 680, 349

\bibitem[{{Ladd} {et~al.}(2005){Ladd}, {Purcell}, {Wong}, \&  {Robertson}}]{Ladd:2005}
{Ladd}, N., {Purcell}, C., {Wong}, T., \& {Robertson}, S. 2005, PASA, 22, 62

\bibitem[{{Lo} {et~al.}(2009){Lo}, {Cunningham}, {Jones}, {Bains}, {Burton},  {Wong}, {Muller}, {Kramer}, {Ossenkopf}, {Henkel}, {Deragopian}, {Donnelly},  \& {Ladd}}]{Lo:2009}
{Lo}, N., {Cunningham}, M.~R., {Jones}, P.~A., {et~al.} 2009, \mnras, 395, 1021

\bibitem[{{L{\'o}pez-Sepulcre} {et~al.}(2011){L{\'o}pez-Sepulcre}, {Walmsley},  {Cesaroni}, {Codella}, {Schuller}, {Bronfman}, {Carey}, {Menten}, {Molinari},  \& {Noriega-Crespo}}]{Lopez-Sepulcre:2011}
{L{\'o}pez-Sepulcre}, A., {Walmsley}, C.~M., {Cesaroni}, R., {et~al.} 2011, \aap, 526, L2+

\bibitem[{{Mardones} {et~al.}(1997){Mardones}, {Myers}, {Tafalla}, {Wilner},  {Bachiller}, \& {Garay}}]{Mardones:1997}
{Mardones}, D., {Myers}, P.~C., {Tafalla}, M., {et~al.} 1997, \apj, 489, 719

\bibitem[{{Peretto} \& {Fuller}(2009)}]{Peretto:2009}
{Peretto}, N. \& {Fuller}, G.~A. 2009, \aap, 505, 405

\bibitem[{{Peretto} {et~al.}(2010){Peretto}, {Fuller}, {Plume}, {Anderson},  {Bally}, {Battersby}, {Beltran}, {Bernard}, {Calzoletti}, {Digiorgio},  {Faustini}, {Kirk}, {Lenfestey}, {Marshall}, {Martin}, {Molinari}, {Montier},  {Motte}, {Ristorcelli}, {Rod{\'o}n}, {Smith}, {Traficante}, {Veneziani},  {Ward-Thompson}, \& {Wilcock}}]{Peretto:2010}
{Peretto}, N., {Fuller}, G.~A., {Plume}, R., {et~al.} 2010, \aap, 518, L98+

\bibitem[{{Pirogov} {et~al.}(2003){Pirogov}, {Zinchenko}, {Caselli},  {Johansson}, \& {Myers}}]{Pirogov:2003}
{Pirogov}, L., {Zinchenko}, I., {Caselli}, P., {Johansson}, L.~E.~B., \&  {Myers}, P.~C. 2003, \aap, 405, 639

\bibitem[{{Purcell} {et~al.}(2006){Purcell}, {Balasubramanyam}, {Burton},  {Walsh}, {Minier}, {Hunt-Cunningham}, {Kedziora-Chudczer}, {Longmore},  {Hill}, {Bains}, {Barnes}, {Busfield}, {Calisse}, {Crighton}, {Curran},  {Davis}, {Dempsey}, {Derragopian}, {Fulton}, {Hidas}, {Hoare}, {Lee}, {Ladd},  {Lumsden}, {Moore}, {Murphy}, {Oudmaijer}, {Pracy}, {Rathborne}, {Robertson},  {Schultz}, {Shobbrook}, {Sparks}, {Storey}, \& {Travouillion}}]{Purcell:2006}
{Purcell}, C.~R., {Balasubramanyam}, R., {Burton}, M.~G., {et~al.} 2006, \mnras, 367, 553

\bibitem[{{Rathborne} {et~al.}(2010){Rathborne}, {Jackson}, {Chambers},  {Stojimirovic}, {Simon}, {Shipman}, \& {Frieswijk}}]{Rathborne:2010}
{Rathborne}, J.~M., {Jackson}, J.~M., {Chambers}, E.~T., {et~al.} 2010, \apj, 715, 310

\bibitem[{{Rawlings} {et~al.}(2004){Rawlings}, {Redman}, {Keto}, \&  {Williams}}]{Rawlings:2004}
{Rawlings}, J.~M.~C., {Redman}, M.~P., {Keto}, E., \& {Williams}, D.~A. 2004,  \mnras, 351, 1054

\bibitem[{{Rieke} {et~al.}(2004){Rieke}, {Young}, {Engelbracht}, {Kelly},  {Low}, {Haller}, {Beeman}, {Gordon}, {Stansberry}, {Misselt}, {Cadien},  {Morrison}, {Rivlis}, {Latter}, {Noriega-Crespo}, {Padgett}, {Stapelfeldt},  {Hines}, {Egami}, {Muzerolle}, {Alonso-Herrero}, {Blaylock}, {Dole}, {Hinz},  {Le Floc'h}, {Papovich}, {P{\'e}rez-Gonz{\'a}lez}, {Smith}, {Su}, {Bennett},  {Frayer}, {Henderson}, {Lu}, {Masci}, {Pesenson}, {Rebull}, {Rho}, {Keene},  {Stolovy}, {Wachter}, {Wheaton}, {Werner}, \& {Richards}}]{Rieke:2004}
{Rieke}, G.~H., {Young}, E.~T., {Engelbracht}, C.~W., {et~al.} 2004, \apjs, 154, 25

\bibitem[{{Schilke} {et~al.}(1997){Schilke}, {Walmsley}, {Pineau des Forets},  \& {Flower}}]{Schilke:1997}
{Schilke}, P., {Walmsley}, C.~M., {Pineau des Forets}, G., \& {Flower}, D.~R.  1997, \aap, 321, 293

\bibitem[{{Schuller} {et~al.}(2009){Schuller}, {Menten}, {Contreras},  {Wyrowski}, {Schilke}, {Bronfman}, {Henning}, {Walmsley}, {Beuther},  {Bontemps}, {Cesaroni}, {Deharveng}, {Garay}, {Herpin}, {Lefloch}, {Linz},  {Mardones}, {Minier}, {Molinari}, {Motte}, {Nyman}, {Reveret}, {Risacher},  {Russeil}, {Schneider}, {Testi}, {Troost}, {Vasyunina}, {Wienen}, {Zavagno},  {Kovacs}, {Kreysa}, {Siringo}, \& {Wei{\ss}}}]{Schuller:2009}
{Schuller}, F., {Menten}, K.~M., {Contreras}, Y., {et~al.} 2009, \aap, 504, 415

\bibitem[{{Shirley} {et~al.}(2003){Shirley}, {Evans}, {Young}, {Knez}, \&  {Jaffe}}]{Shirley:2003}
{Shirley}, Y.~L., {Evans}, II, N.~J., {Young}, K.~E., {Knez}, C., \& {Jaffe},  D.~T. 2003, \apjs, 149, 375

\bibitem[{{Shukla} {et~al.}(2004){Shukla}, {Yun}, \& {Scoville}}]{Shukla:2004}
{Shukla}, H., {Yun}, M.~S., \& {Scoville}, N.~Z. 2004, \apj, 616, 231

\bibitem[{{Simon} {et~al.}(2006){Simon}, {Jackson}, {Rathborne}, \&  {Chambers}}]{Simon:2006a}
{Simon}, R., {Jackson}, J.~M., {Rathborne}, J.~M., \& {Chambers}, E.~T. 2006,  \apj, 639, 227

\bibitem[{{Turner} \& {Thaddeus}(1977)}]{Turner:1977}
{Turner}, B.~E. \& {Thaddeus}, P. 1977, \apj, 211, 755

\bibitem[{{Walsh} \& {Burton}(2006)}]{Walsh:2006}
{Walsh}, A.~J. \& {Burton}, M.~G. 2006, \mnras, 365, 321

\bibitem[{{Walsh} {et~al.}(2008){Walsh}, {Lo}, {Burton}, {White}, {Purcell},  {Longmore}, {Phillips}, \& {Brooks}}]{Walsh:2008}
{Walsh}, A.~J., {Lo}, N., {Burton}, M.~G., {et~al.} 2008, PASA, 25, 105

\bibitem[{{Watson} {et~al.}(2008){Watson}, {Povich}, {Churchwell}, {Babler},  {Chunev}, {Hoare}, {Indebetouw}, {Meade}, {Robitaille}, \&  {Whitney}}]{Watson:2008}
{Watson}, C., {Povich}, M.~S., {Churchwell}, E.~B., {et~al.} 2008, \apj, 681, 1341

\bibitem[{{Williams} {et~al.}(2000){Williams}, {Blitz}, \&  {McKee}}]{Williams:2000}
{Williams}, J.~P., {Blitz}, L., \& {McKee}, C.~F. 2000, Protostars and Planets  IV, 97

\bibitem[{{Wu} {et~al.}(2010){Wu}, {Evans}, {Shirley}, \& {Knez}}]{Wu:2010}
{Wu}, J., {Evans}, N.~J., {Shirley}, Y.~L., \& {Knez}, C. 2010, \apjs, 188, 313

\end{thebibliography}

\clearpage
\LongTables
\begin{deluxetable}{cccccccc}
\tablecaption{MALT90 Pilot Sources}
	\tabletypesize{\scriptsize}

	\tablehead{\colhead{Name} & \colhead{Glon} & \colhead{Glat} & \colhead{RA J2000} & \colhead{Dec J2000}  &\colhead{Catalog} & \colhead{Velocity} & \colhead{T$_{rms}$} \\
	& \colhead{[degrees]} & \colhead{[degrees]} & \colhead{[HH:MM:SS]} & \colhead{[DD:MM:SS]} & &  \colhead{LSR [km s$^{-1}$]} & \colhead{[K]} }
	\startdata
G263.620$-$00.530	 & 	263.620	 & 	$-$0.530	 & 	08:45:36.003	 & 	-43:51:01.99	 & 	IRAS	 & 	+4.72	 & 	0.282\\
G264.322$-$00.184	 & 	264.322	 & 	$-$0.184	 & 	08:49:32.995	 & 	-44:10:46.19	 & 	IRAS	 & 	+9.07	 & 	0.364\\
G285.259$-$00.049	 & 	285.259	 & 	$-$0.049	 & 	10:31:28.383	 & 	-58:02:07.23	 & 	IRAS	 & 	+2.14	 & 	0.199\\
G287.814$-$00.816	 & 	287.814	 & 	$-$0.816	 & 	10:45:53.955	 & 	-59:57:01.87	 & 	IRAS	 & 	-14.96	 & 	0.195\\
G300.958$+$00.902	 & 	300.958	 & 	$+$0.902	 & 	12:34:39.728	 & 	-61:54:19.57	 & 	MIPSGAL	 & 	-41.67	 & 	0.209\\
G300.960$-$00.702	 & 	300.960	 & 	$-$0.702	 & 	12:33:44.636	 & 	-63:30:21.53	 & 	GLIMPSE-Bright	 & 	\nodata	 & 	0.208\\
G300.968$+$01.145	 & 	300.968	 & 	$+$1.145	 & 	12:34:52.771	 & 	-61:39:48.93	 & 	ATLASGAL	 & 	-42.96	 & 	0.197\\
G301.991$-$00.704	 & 	301.991	 & 	$-$0.704	 & 	12:42:59.051	 & 	-63:33:37.09	 & 	GLIMPSE-Bright	 & 	\nodata	 & 	0.177\\
\enddata
\label{basicdata}
\tablecomments{Central source velocities are determined by hand-examination of the four main lines for each source. Where two velocity components are seen, the stronger is listed first. T$_{rms}$ gives the noise per channel in the spectrum as measured at the central position in the map.}
\end{deluxetable}
\clearpage

\clearpage

\begin{deluxetable}{ccccccccccc}
\tablecaption{Properties of Main Lines at Position of Maximum Integrated Intensity}
	\tabletypesize{\scriptsize}

	\tablehead{\colhead{Name} & \multicolumn{4}{c}{Maximum Integrated Intensity [K km s$^{-1}$]} & \multicolumn{4}{c}{Offset from Map Center\tablenotemark{a} [9\arcsec\ pixels]} & \multicolumn{2}{c}{FWHM [km s$^{-1}$]} \\
	& \colhead{\nthp} & \colhead{\hnc} & \colhead{\hcop} & \colhead{\hcn} & \colhead{\nthp} & \colhead{\hnc} & \colhead{\hcop} & \colhead{\hcn} & \colhead{\hnc} & \colhead{\hcop}
	}
	\startdata
G263.620$-$00.530	 & 	2.47$\pm$0.42	 & 	2.95$\pm$0.38	 & 	5.09$\pm$0.39	 & 	5.18$\pm$0.38	 & 	(9, -2)	 & 	(-4, -5)	 & 	(-5, -5)	 & 	(-4, -4)	 & 	6.33$\pm$1.59	 & 	5.59$\pm$1.09\\
G264.322$-$00.184	 & 	3.58$\pm$0.57	 & 	3.15$\pm$0.49	 & 	3.82$\pm$0.56	 & 	4.16$\pm$0.46	 & 	(-9, -2)	 & 	(-7, -3)	 & 	(-9, -4)	 & 	(-4, -3)	 & 	6.18$\pm$1.90	 & 	3.95$\pm$4.01\\
G285.259$-$00.049	 & 	4.03$\pm$0.25	 & 	4.34$\pm$0.26	 & 	13.07$\pm$0.27	 & 	12.42$\pm$0.28	 & 	(0, -8)	 & 	(-1, -8)	 & 	(-3, -1)	 & 	(-3, -1)	 & 	0.39$\pm$13.50	 & 	5.00$\pm$0.37\\
G287.814$-$00.816	 & 	1.65$\pm$0.25	 & 	2.96$\pm$0.25	 & 	7.71$\pm$0.26	 & 	9.12$\pm$0.27	 & 	(1, 4)	 & 	(-1, 2)	 & 	(-1, 2)	 & 	(-1, 2)	 & 	3.53$\pm$1.99	 & 	0.42$\pm$7.32\\
G300.958$+$00.902	 & 	3.14$\pm$0.26	 & 	3.20$\pm$0.28	 & 	4.39$\pm$0.27	 & 	3.22$\pm$0.28	 & 	(3, -5)	 & 	(7, 3)	 & 	(8, 3)	 & 	(6, 1)	 & 	5.62$\pm$1.21	 & 	6.00$\pm$0.79\\
G300.960$-$00.702	 & 	$<$1.38	 & 	$<$1.38	 & 	$<$1.38	 & 	$<$1.38	 & 	\nodata       	 & 	\nodata       	 & 	\nodata       	 & 	\nodata       	 & 	\nodata       	 & 	\nodata       \\
G300.968$+$01.145	 & 	5.11$\pm$0.24	 & 	5.83$\pm$0.26	 & 	12.62$\pm$0.26	 & 	16.05$\pm$0.27	 & 	(1, -1)	 & 	(0, 0)	 & 	(0, 0)	 & 	(-1, 0)	 & 	2.20$\pm$1.80	 & 	4.76$\pm$0.35\\
G301.991$-$00.704	 & 	$<$1.17	 & 	$<$1.17	 & 	$<$1.17	 & 	$<$1.17	 & 	\nodata       	 & 	\nodata       	 & 	\nodata       	 & 	\nodata       	 & 	\nodata       	 & 	\nodata       \\
\enddata
\label{mainlines}
\tablecomments{Results are listed for lines with $5\sigma$ detections in the narrow ($\pm 2.25$ km s$^{-1}$) moment maps, but these measurements report the moments calculated in a range of ($\pm 8.25$ km s$^{-1}$). Not all integrated intensity measurements at $5\sigma$ results in this broader velocity range.}
\tablenotetext{a}{Offsets are (x,y) offsets in units of pixels, where each pixel is 9\arcsec. Offsets are relative to the targeted center of the map, which is determined differently for the different input surveys.}
\tablenotetext{b}{\nthp\ is outside the spectral coverage for sources with V$_{LSR} < -100 $km s$^{-1}$}
\end{deluxetable}

\clearpage

\begin{deluxetable}{cccccccccccc}
\tablecaption{Robust (5$\sigma$) Detections of Lines}
	\tabletypesize{\scriptsize}
	\tablehead{\colhead{Name} & \colhead{\nthp} & \colhead{\tcs} 
	& \colhead{\hnc} & \colhead{\hcop} & \colhead{\hcn}
	& \colhead{\cth} & \colhead{\sio} & \colhead{\htcop} & \colhead{\htcn}
	}
		\startdata
G263.620$-$00.530	 & 	1	 & 	0	 & 	1	 & 	1	 & 	1	 & 	1	 & 	0	 & 	0	 & 	0 \\
G264.322$-$00.184	 & 	1	 & 	0	 & 	1	 & 	1	 & 	1	 & 	0	 & 	0	 & 	0	 & 	0 \\
G285.259$-$00.049	 & 	1	 & 	0	 & 	1	 & 	1	 & 	1	 & 	1	 & 	0	 & 	0	 & 	0 \\
G287.814$-$00.816	 & 	1	 & 	0	 & 	1	 & 	1	 & 	1	 & 	1	 & 	0	 & 	0	 & 	0 \\
G300.958$+$00.902	 & 	1	 & 	0	 & 	1	 & 	1	 & 	1	 & 	1	 & 	0	 & 	1	 & 	0 \\
G300.960$-$00.702	 & 	0	 & 	0	 & 	0	 & 	0	 & 	0	 & 	0	 & 	0	 & 	0	 & 	0 \\
G300.968$+$01.145	 & 	1	 & 	0	 & 	1	 & 	1	 & 	1	 & 	1	 & 	0	 & 	1	 & 	1 \\
G301.991$-$00.704	 & 	0	 & 	0	 & 	0	 & 	0	 & 	0	 & 	0	 & 	0	 & 	0	 & 	0 \\
\enddata
\label{detectiontable}

\tablecomments{Robust detections correspond to 5$\sigma$ integrated intensity detections excluding the 3 pixels (27\arcsec) on the edge of each map. Detection statistics are not shown for \hfa, \chtcn, \hctn, \tcts, \hctccn, \hncofot\ or \hncofzf, which have no robust detections. One corresponds to a robust detection, zero indicates no robust detection.}

\end{deluxetable}

\end{document}